\documentclass[twocolumn,english,aps]{revtex4}

\usepackage{amstext}
\usepackage{graphicx}
\usepackage{amssymb}
\usepackage{setspace}
\usepackage{babel}
\usepackage{cases}
\usepackage{hyperref}
\usepackage{subfigure}
\usepackage{verbatim}
\usepackage{soul}
\begin{document}

\title{Quantum simulation of 2d topological physics using orbital-angular-
momentum-carrying photons in a 1d array of cavities}

\author{
 Xi-Wang Luo,
 Xingxiang Zhou \footnote{email: xizhou@ustc.edu.cn},
 Chuan-Feng Li \footnote{email: cfli@ustc.edu.cn},
 Jin-Shi Xu,
 Guang-Can Guo,
 and Zheng-Wei Zhou \footnote{email: zwzhou@ustc.edu.cn}
}
\affiliation{
$^1$Key Laboratory of Quantum Information, University of Science and
Technology of China, Hefei, Anhui 230026, China\\
$^2$Synergetic Innovation Center of Quantum Information and Quantum Physics,
University of Science and Technology of China, Hefei, Anhui 230026, China
}

\newpage

\begin{abstract}
Orbital angular momentum (OAM) of light is a fundamental optical degree of
freedom that has recently motivated much exciting research in diverse fields
ranging from optical communication to quantum information. We show for the
first time that it is also a unique and valuable resource for quantum
simulation, by demonstrating theoretically how \emph{2d} topological
physics can be simulated in a \emph{1d} array of optical cavities using
OAM-carrying photons. Remarkably, this newly discovered application of OAM
states not only reduces required physical resources but also increases feasible
scale of simulation. By showing how important topics such as edge-state
transport and topological phase transition can be studied in a small simulator
with just a few cavities ready for immediate experimental exploration, we
demonstrate the prospect of photonic OAM for quantum simulation
which can have a significant impact on the research of topological physics.

\end{abstract}

\maketitle

As a relatively under-exploited optical degree of freedom, OAM of photons has
attracted much research interest lately. Beams of OAM-carrying photons have an
azimuthal phase dependence in the form $e^{il\varphi}$ where the OAM quantum
number $l$ can take any integer value \cite{allen1992orbital}. These photon
modes, which arise in the natural
solutions of the paraxial wave equation in cylindrical coordinates
\cite{yariv2007photonics}, can be manipulated and measured
with high precision \cite{mair2001entanglement,
lee2004optical, malik2014direct, yao2011orbital}. Because of the
unlimited range of the angular
momentum, OAM-carrying photons are recognized as a unique asset in many studies.
On the application side, they are used to enable high-capacity optical
communication \cite{barreiro2008beating, wang2012terabit} and versatile optical
tweezers \cite{padgett2011tweezers}. In fundamental research, they have played
important roles in quantum information and quantum foundation
\cite{nagali2009quantum, fickler2014interface, ding2013single,
nicolas2014a, leach2010quantum, molina2007twisted, yao2011orbital}.
Though the study of OAM states used to be limited to low angular
momenta, there has been tremendous advance lately motivated by their great
potential. This is highlighted by the remarkable recent demonstration of quantum
entanglement involving angular momenta as high as hundreds
\cite{fickler2012quantum,krenna14}.

In this work, we show theoretically that OAM of photons are also very
useful for nontrivial quantum simulation, a potential that has not been realized
and considered before. Specifically, we demonstrate how they can be used to
simulate a broad range of topological physics which are at the heart of a group
of extraordinary quantum phenomena that arise in \emph{2d} systems subject to
external gauge
fields. These include the likes of integer \cite{klitzing1980new} and
fractional \cite{tsui1982two} quantum Hall effect and quantum spin Hall
effect \cite{konig2007quantum}, which are characterized by
exotic properties such as quantized conductance and edge-state transport.
Topological effects are often difficult to investigate due to
stringent experimental conditions involved, and some theoretical
predictions remain challenging to observe \cite{hasan2010colloquium,
konig2007quantum}. To overcome this difficulty, various simulation schemes
based on different physical platforms such as
ultra-cold atoms \cite{dalibard2011colloquium, NJP.12.033041, NJP.14.015007} and
photons \cite{PhysRevLett.100.013904, wang2009observation, cho2008fractional,
umucalilar2011artificial, hafezi2011robust, fang2012realizing,
khanikaev2012photonic, hafezi2013imaging, rechtsman2013photonic, PhysRevLett.109.106402, Lu2014} have been suggested recently. Not surprisingly, central
to most simulation schemes is a \emph{2d} architecture for the simulator. Many
of them are still very demanding, requiring limit-pushing experimental
conditions or advanced new technologies.

In contrast to other proposals \cite{PhysRevLett.100.013904, wang2009observation, cho2008fractional,
umucalilar2011artificial, hafezi2011robust, fang2012realizing,
khanikaev2012photonic, hafezi2013imaging, rechtsman2013photonic}, our system has
a \emph{1d} structure which does not need to be large in scale, thus greatly
reducing the complexity of the system. Remarkably, feasible scale of simulation is
increased despite the simplified system, and it is so versatile that the
effect of arbitrary Abelian and non-Abelian gauge fields can be studied using
standard linear optics devices only, with no restriction on the form of the
gauge fields \cite{hafezi2011robust, fang2012realizing, rechtsman2013photonic}
and no need for specially designed meta-material \cite{khanikaev2012photonic} or
photonic crystal \cite{rechtsman2013photonic}. It then allows to investigate
important topological problems under intense pursuit such as non-Abelian gauge
field induced phase transition between a photonic normal and topological
insulator. Further, we can easily probe the topological properties
of our system by measuring the photon transmission coefficients which are shown
to have deep connections to the essential topological invariants of the system.
All this is possible because of the inherent properties of the OAM of
photons, whose power and
potential for quantum simulation is just recognized and can be unleashed
readily.

\section*{Results}

\subsubsection*{The 1d array of cavities}

\begin{figure}
\includegraphics[width=1.0\linewidth]{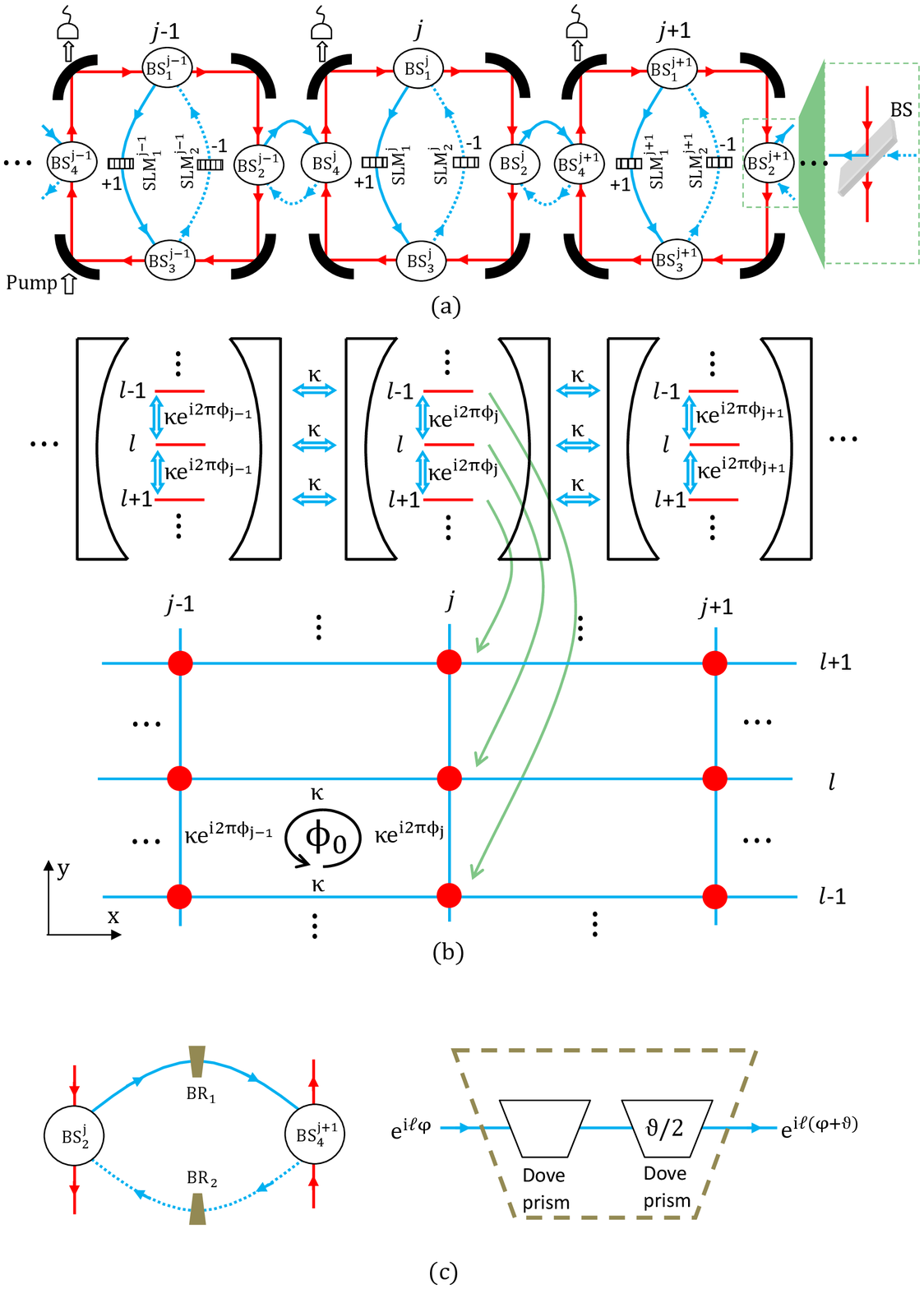}
\caption{A \emph{1d} array of degenerate cavities for simulating a \emph{2d}
rectangular lattice in a magnetic field. (a) The optical design
for simulating $\mathcal{H}_1$. Each main cavity has an auxiliary
cavity consisting of two BSs (BS$_1^j$ and BS$_3^j$) and two SLMs (SLM$_1^j$ and
SLM$_2^j$). There is also a coupling cavity (made of BS$_2^j$ and
BS$_4^{j+1}$) between adjacent main cavities (It can be replaced with a simple
BS to reduce the number of optical elements in experiments).
The length of both the auxiliary and coupling cavity is chosen for destructive
interference, and most light remains in the main cavity. The cavities at
the two ends of the array can be coupled to realize periodic boundary
condition, or uncoupled for open boundary condition.
(b) Mapping of the \emph{1d} simulator array in (a) to a \emph{2d} rectangular
lattice in a magnetic field.
(c) The coupling cavity (left) for simulating $\mathcal{H}_5$ and the optical
design (right) for the beam rotators BR$_1$ and BR$_2$ with opposite
rotation angles $\pm\vartheta=\pm 2\pi\phi_0$. The main cavity and auxiliary
cavity require no modification, except that the phase difference between the
arms containing the SLMs is set to 0.
}
\label{fig:simulator_Abelian}
\end{figure}

Shown in Fig. \ref{fig:simulator_Abelian} (a) is our simulation system. It
consists of an array of $N$ nominally identical cavities that are coupled along
the $x$ direction. The system size, $N$,
does not need to be large; we will show that even a simulator with just a few
cavities is sufficient to demonstrate topological effects. The building blocks
are degenerate cavities \cite{arnaud1969degenerate,
chalopin2010frequency}, which have appropriate optical design such that they
can support photon modes with different OAM (Supplementary Note 1). In
each cavity, we make use of clockwise circulating OAM-carrying
photons and denote their annihilation operator $\hat{a}_{j,l}$, where $j$
($0\leq j \leq N-1$) is the index of the cavity in the array and $l$ is the OAM
number of the photon mode. To manipulate the OAM state of photons,
for each cavity we introduce an auxiliary cavity consisting of two beam
splitters (BSs) and two spatial light modulators (SLMs). The BSs divert a
portion of the light
in the main cavity toward the
SLMs and merge it back. When propagating between the BSs, photons can
accumulate a phase. The SLMs, which can be simple spiral phase plates with
very low loss \cite{PhysRevLett.95.240501, oemrawsingh2004production},
change the OAM of photons by $\pm 1$.

As depicted in Fig. \ref{fig:simulator_Abelian} (b), by associating the OAM
number of the photon in a cavity with the site index number along the $y$
direction of an (imaginary)
lattice, we can conceptually map our \emph{1d} array of cavities to a \emph{2d}
rectangular lattice system. In Fig. \ref{fig:simulator_Abelian} (a),
the BSs and SLMs of the auxiliary cavity change the OAM of a portion of
the light in the main cavity by
$\pm 1$, and this corresponds to hopping of a particle on the lattice site in
Fig. \ref{fig:simulator_Abelian} (b) along the $y$ direction to its
neighboring sites with a
probability determined by the reflectivity of the BSs. In this hopping process,
the particle can also acquire an experimentally controllable phase determined by
the imbalance between the optical paths from BS$_1^j$ to BS$_3^j$ and backwards.
As shown in the Supplementary Note 2, the Hamiltonian of the simulated
system is
\begin{eqnarray}
\mathcal{H}_1=-\kappa\sum_{j,l} \left(e^{i2\pi\phi_j}
\hat{a}_{j,l+1}^{\dagger}\hat{a}_{j,l}+\hat{a}_{j+1,l}^{\dagger}\hat{a}_{j,l}+h.c.\right), \nonumber
\label{eq:H_Abelian}
\end{eqnarray}
where $\kappa$ is the transition rate between different OAM states, chosen to
be the same with the coupling rate between neighboring cavities, and
$2\pi\phi_j$ is the phase acquired by the photon in
the $j$-th cavity when it travels between the BSs in the auxiliary cavity. If we
set up the system such that $\phi_j$ is linearly dependent on the cavity index
$j$, $\phi_j = j\phi_0$, then $\mathcal{H}_1$  describes a tight-binding model
of charged particle in a \emph{2d} lattice subject to a uniform magnetic field
with $\phi_0$ quanta of flux per plaquette \cite{hofstadter1976energy}.

\begin{figure}
\includegraphics[width=0.8\linewidth]{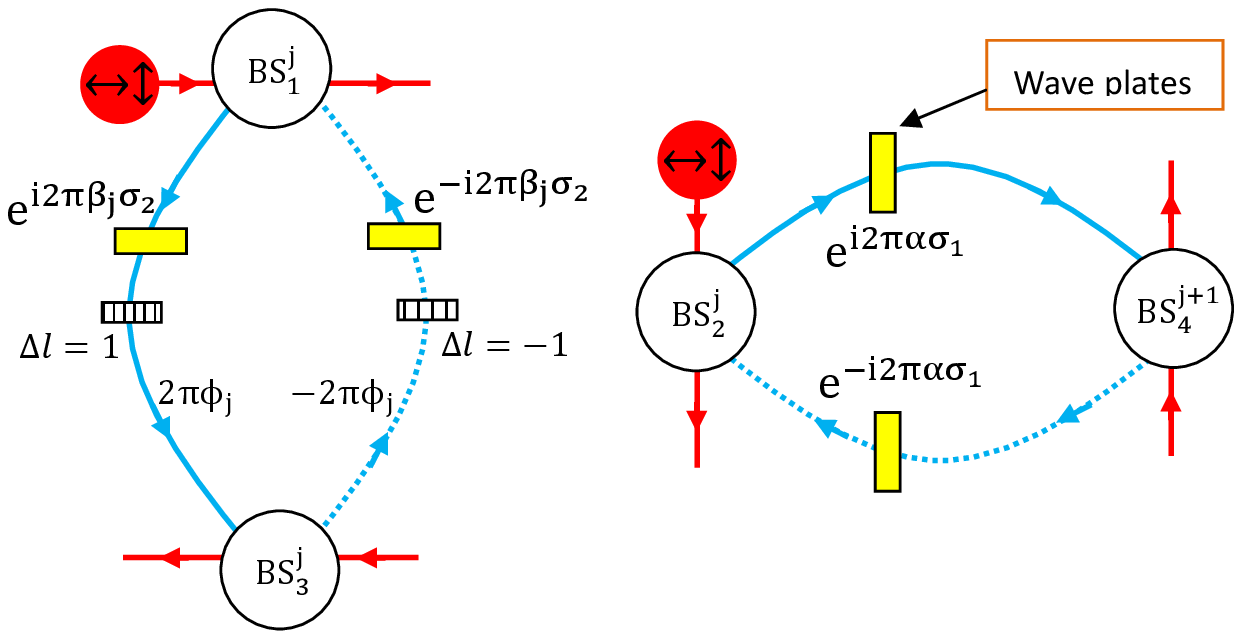}
\caption{Design of the auxiliary and coupling cavity for simulating
$\mathcal{H}_2$ with polarized photons. The main
cavity requires no modification. The birefringent waveplates can induce
different phase delays for the two polarizations and cause transitions between
them when their optical axes are properly aligned (Supplementary Note 3). Their
effect is described by the Jones matrices $e^{\pm i2\pi\alpha \sigma_1}$ and
$e^{\pm i2\pi\beta \sigma_2}$, where the two polarizations are represented by
the spin up and down and
the Pauli matrices $\sigma_1=\vec{\sigma} \cdot\mathbf{n}_1$,
$\sigma_2=\vec{\sigma} \cdot\mathbf{n}_2$ with $\mathbf{n}_1$ and
$\mathbf{n}_2$ two unit vectors determined by the design of the waveplates.
}
\label{fig:simulator_nonAbelian}
\end{figure}

Therefore, by representing a spatial degree of freedom with the OAM states of
photons, we can study a \emph{2d} system with a \emph{1d} simulator, greatly
reducing the physical resources required for the simulation. Unlike in
earlier \emph{1d} optical simulator \cite{PhysRevLett.109.106402}, our system
performs a full and genuine \emph{2d} simulation, rather than simulate the
\emph{1d} behavior of the system at a fixed Bloch momentum in the other
direction. Meanwhile, in comparison with a \emph{2d} array of coupled
cavities, the size of the \emph{2d} lattice that can be simulated is
dramatically increased along
the $y$ direction. This is due to the fact that, unlike in an atomic
system \cite{PhysRevLett.112.043001} where only a small number of atomic states
are available for the simulation, there is no upper limit for the OAM of photons
in theory. In reality, it is limited by practical factors such as the size of
the optical
elements and can be made very large in a proper design.
In contrast, the feasible size in the $y$ direction for a \emph{2d} cavity
array would be much smaller, because nonuniformity of the cavities and local
disturbances will make photons quickly lose coherence after traveling
through a few cavities.
This remarkable combination of reduced physical resources and increased
scale of simulation makes our system very promising.
Also, our system can be easily modified to support more demanding simulations
by making use of additional degrees of freedom of photons. For instance, we
can simulate the quantum spin Hall (QSH) effect \cite{kane2005z} in non-Abelian
gauge fields \cite{bermudez10, mazza12} by using the horizontal and
vertical polarizations of polarized photons to represent the up and down state
($s=\pm1$) of a spin. By using birefringent waveplates whose optical
axes are properly aligned with respect to the horizontal and vertical
polarizations, we can assign different phases to the two polarizations and cause
transitions between them when they pass the waveplates (see Supplementary Note 3
for details). We can then manipulate the polarization states of the photon to
mimick the spin flips and spin-dependent phase delays caused by non-Abelian
gauge fields, as illustrated in Fig. \ref{fig:simulator_nonAbelian}.
The simulated Hamiltonian is (Supplementary Note 3)
\begin{eqnarray}
\mathcal{H}_2=&-&\kappa\sum_{j,l}\left(\mathbf{\hat{a}}_{j,l+1}^{\dagger}e^{
i2\pi\hat{\theta}_y}\mathbf{\hat{a}}
_{j,l}
+\mathbf{\hat{a}}_{j+1,l}^{\dagger}e^{i2\pi\hat{\theta}_x}\mathbf{\hat{a}}_{j,l}
+h.c.\right)\nonumber
\\
&+&\sum_{j,l}\lambda_j\mathbf{\hat{a}}_{j,l}^{\dagger}\mathbf{\hat{a}}_{j,l},
\label{eq:H_nonAbelian}
\end{eqnarray}
where $\mathbf{\hat{a}}^{\dagger}_{j,l} =
({a}^{\dagger}_{j,l,\leftrightarrow},{a}^{\dagger}_{j,l,\updownarrow})$ is a
two-component (the horizontal
and vertical polarization) photon creation operator,
and $\lambda_j$ is an effective on-site energy.
The tunneling phases, which correspond to the potentials of the associated
gauge fields \cite{dalibard2011colloquium}, are given by
\begin{eqnarray}
\hat{\theta}_x=\alpha\hat\sigma_1, \hat{\theta}_y=\phi_j+\beta_j\hat\sigma_2,
\label{eq:H_param}
\end{eqnarray}
where $\phi_j$ is the spin-independent part of the phase, and
$\alpha$, $\beta_j$, $\hat\sigma_1$ and $\hat\sigma_2$ are determined by the
Jones matrices \cite{yariv2007photonics} of the waveplates as shown in
Fig. \ref{fig:simulator_nonAbelian}.
By selecting appropriate waveplates and manipulating the polarization of the
photon accordingly, we can engineer non-commuting tunneling phases
$\hat{\theta}_x$ and $\hat{\theta}_y$, and thus simulate the effect of an
arbitrary non-Abelian gauge field.

\subsubsection*{Probing scheme}
Since we represent a spatial degree of freedom with OAM states of
photons, the measurement of our system involves manipulation and detection
of the OAM states. Specifically, we pump the $j_i$-th cavity using a probing
light with a definitive OAM $l_i$
and measure in the steady state how much ends up in the OAM mode $l_o$ in the
$j_o$-th cavity by leaking
a small amount of light out of each cavity, as shown in Fig.
\ref{fig:simulator_Abelian} (a). It is determined by the
transmission coefficient \cite{PhysRevB.59.15882}
(Supplementary Note 4)
\begin{equation}
T_{j_i, l_i}^{j_o, l_o}(\omega)=-i\gamma\bigg\langle j_o, l_o\bigg|
\frac{1}{\omega-\mathcal{H}_{SYS}+i\gamma/2}\bigg|j_i, l_i, \bigg\rangle,
\label{eq:T}
\end{equation}
where $\omega$ is the detuning of the probing light from the cavity frequency,
$\gamma$ is the photon loss of the system, and $\mathcal{H}_{SYS}$ is
the simulated Hamiltonian. When
non-Abelian gauge fields are concerned, the polarization indexes $s_i$ and
$s_o$ should also be included for the input and output modes.

Generation and detection of OAM-carrying photons can be accomplished very
reliably \cite{mair2001entanglement, yao2011orbital}.
By a coherent measurement, we can determine both the amplitude and phase of
$T_{j_i, l_i}^{j_o, l_o}(\omega)$.
Thanks to the $1d$ structure
of our system and the use of OAM states, we can perform this measurement between
any pair of $(j_i, l_i)$ and $(j_o, l_o)$, equivalent to measuring the
transmission coefficient between any pair of sites in the simulated $2d$
lattice. Such powerful probing capability is key to the demonstration of various
topological effects in our system.

Feasible measurement and clear demonstration of topological properties
is the topic of many recent studies \cite{hasan2010colloquium, khanikaev2012photonic, hafezi2013imaging, atala2013direct, goldman2013direct, wang2013proposal} since
generally speaking it is a very challenging task. Remarkably, in our system it
is straightforward and requires no more than measuring the photon transmission
coefficient in equation (\ref{eq:T}). As we will show, there is a deep
connection between the photon transmission coefficient and the essential
topological invariants which can be exploited to demonstrate topological
behavior in optical systems.

\subsubsection*{System spectrum and density of states}
\begin{figure}
\includegraphics[width=1.0\linewidth]{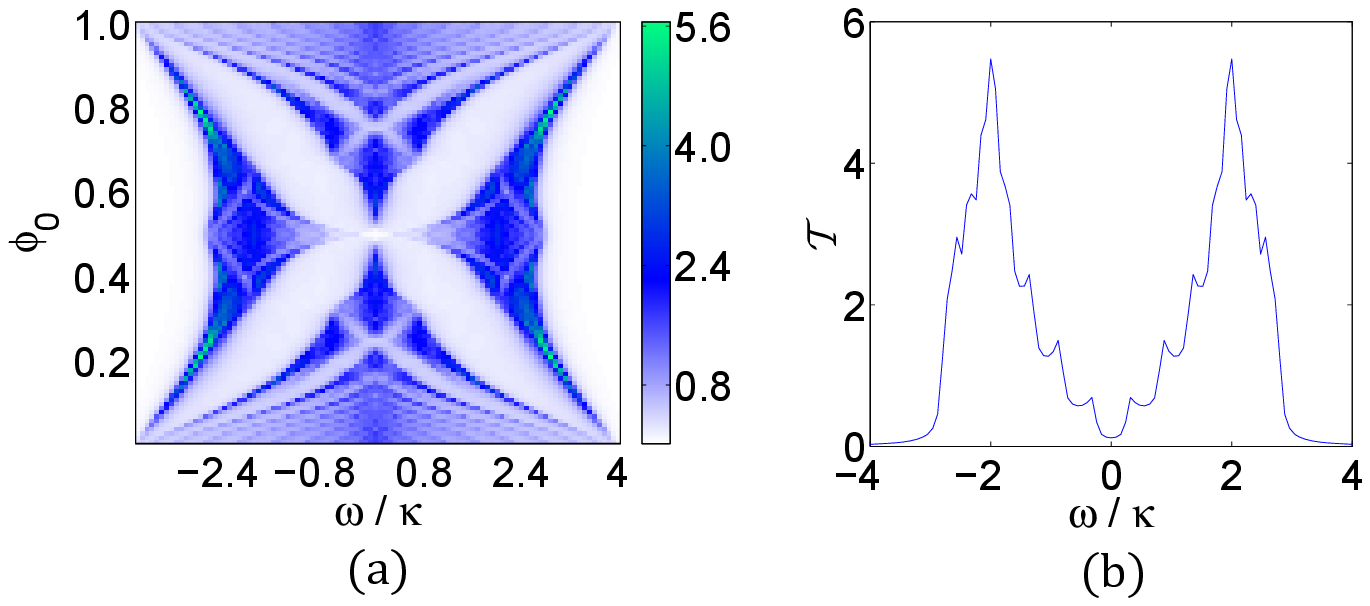}
\caption{Simulation of photon transmission spectroscopy. (a)
Calculated transmission spectra
$\sum_{j_i=0}^{N-1} \mathcal{T}_{j_i,0}$ of
$\mathcal{H}_1$ under different values of magnetic flux $\phi_0$.
Since it is possible to measure the
transmission coefficient between every pair
of lattice sites, we add transmissions to all output channels to obtain a strong
signal and increase the sensitivity of the measurement.
(b) Calculated transmission spectrum
$\mathcal{T}=\sum_{j_i,s_i}\mathcal{T}_{j_i,0,s_i}$ of
$\mathcal{H}_3$, where $\mathcal{T}_{j_i,0,s_i}
=\sum_{j_o,l_o,s_o}|T_{j_i, 0, s_i}^{j_o,l_o, s_o}(\omega)|^2$. In both
(a) and (b), the size of the simulator $N=10$. The OAM number of the photon
included in the calculation is $l\in[-50,50]$. Open and periodic boundary
conditions are used in the $x$ and $y$ direction. The photon loss
$\gamma=0.1\kappa$.
}
\label{fig:spectrum}
\end{figure}

As can be seen in equation (\ref{eq:T}), $T_{j_i, l_i}^{j_o, l_o}(\omega)$ is
sensitive to the energy mismatch between the frequency of the probing light and
the energy of the system.
Because of this, we can study the system's spectrum by measuring the
transmission coefficient
\begin{eqnarray}
\mathcal{T}_{j_i, l_i}(\omega)= \sum_{j_o,l_o}
\mathcal{T}_{j_i,l_i}^{j_o,l_o}(\omega)
=
\sum_{j_o,l_o}|T_{j_i,
l_i}^{j_o,l_o}(\omega)|^2 \nonumber
\label{eq:T_spectrum}
\end{eqnarray}
as a function of the frequency of the probing light, where
$\mathcal{T}_{j_i,l_i}^{j_o, l_o} =|T_{j_i,l_i}^{j_o,l_o}|^2$. For a system in an Abelian
gauge field described by $\mathcal{H}_1$, we calculate and
plot in Fig. \ref{fig:spectrum} (a) the system spectrum which
is the well-known Hofstadter butterfly
\cite{hofstadter1976energy}. We see that the main characteristics of the system
spectrum are clearly identifiable even in a small simulator with just a few
cavities.

The transmission spectroscopy is also very valuable for studying physics
associated with a non-Abelian gauge field. As an example, in equation
(\ref{eq:H_param}), if we choose $\hat{\sigma}_1=\sigma_y$,
$\hat\sigma_2=\sigma_x$, $\beta_j=\beta=\alpha=\frac{1}{4}$, $\lambda_j=0$, and
$\phi_j=j\phi_0=0$, we get the \emph{2d} Dirac's Hamiltonian on a
lattice
\cite{boada2011dirac}
\begin{eqnarray}
\mathcal{H}_3=-i\kappa\sum_{j,l}\left(\mathbf{\hat{a}}_{j,l+1}^{\dagger}e^{
i2\pi j\phi_0}\sigma_x\mathbf{\hat{a}}_{j,l}
+\mathbf{\hat{a}}_{j+1,l}^{\dagger}\sigma_y\mathbf{\hat{a}}_{j,l} \right) +h.c., \nonumber
\label{eq:H_Dirac}
\end{eqnarray}
which is a topic of intense research because of its importance
for understanding the properties of graphene and other exotic systems
\cite{goldman2009ultracold, RevModPhys.81.109, NJP.12.033041, NJP.14.015007}.
Characteristic of $\mathcal{H}_3$ are four conical singularities at the Dirac
points \cite{goldman2009ultracold} in the spectrum which give rise to massless
relativistic particles.
As the energy deviates from the Dirac points,
the change of the dispersion relation from relativistic to
non-relativistic is revealed by the Van Hove singularities (VHS) in the density
of states (DOS). When the decay rate $\gamma$ is small, the DOS can be
inferred from the photon transmission spectrum which is shown in Fig.
\ref{fig:spectrum} (b).
The Dirac point at $\omega=0$ and two VHS near $\omega=\pm 2 \kappa$
are observed, confirming Dirac physics related behavior in the system.

\subsubsection*{Edge states and topological protection}
\begin{figure}
\includegraphics[width=1.0\linewidth]{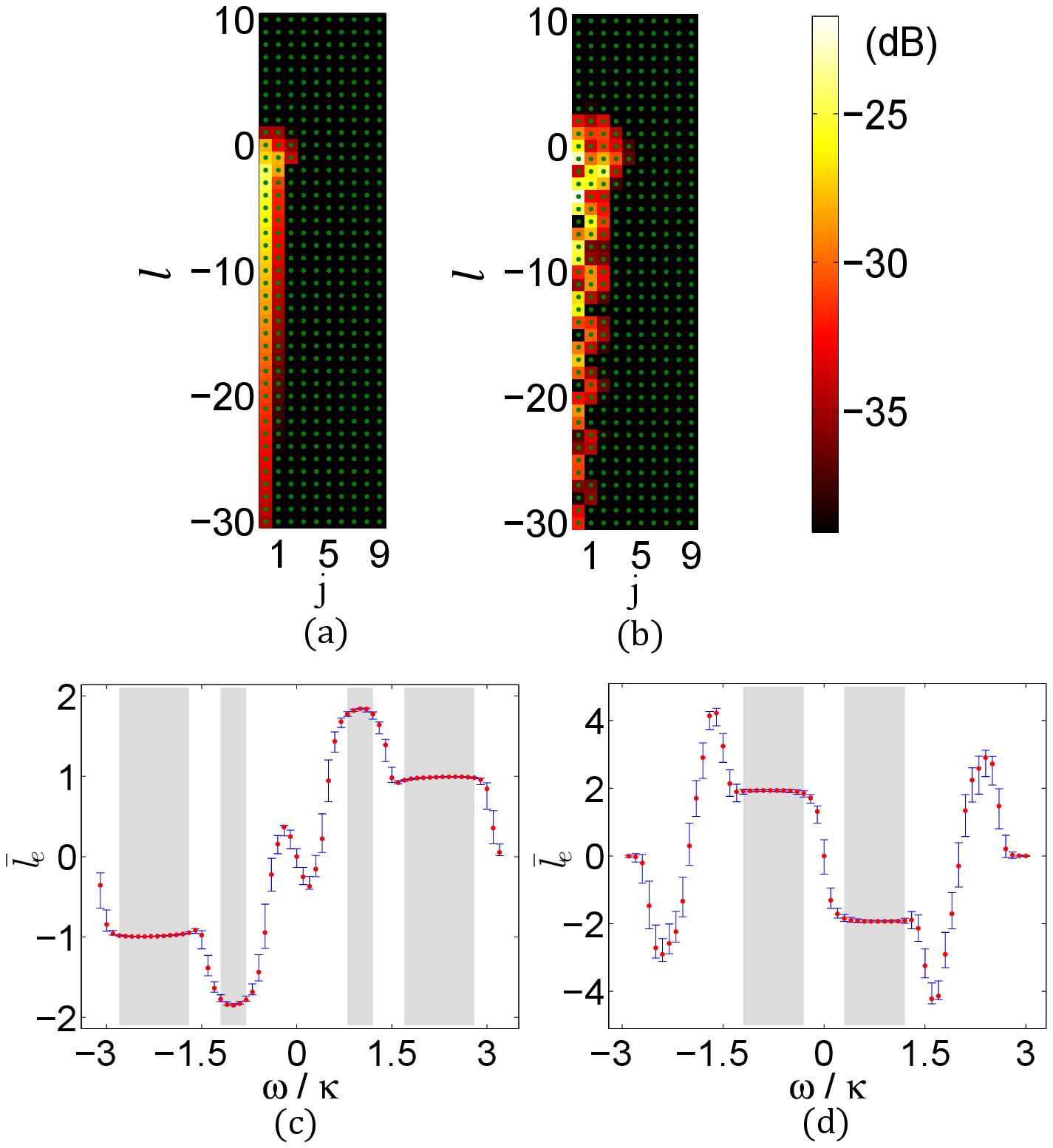}
\caption{Simulation of edge-state transport. (a) Calculated
photon transmission $\mathcal{T}_{0,0}^{j,l}
\equiv|T_{0,0}^{j,l}|^2$ for $\mathcal{H}_1$ with $\phi_0=1/6$.
The frequency of the probing light,
$\omega=-2.2\kappa$, is located in the middle of the first band gap. There is 1
edge mode in this large band gap.
(b) Calculated photon transmission when the probing light frequency
$\omega=-1.0\kappa$
is located in the smaller second gap. Two edge modes are present and
interference patterns due to their phase velocity mismatch are observed.
(c) $\bar{l}_e$ (red dots) and its standard deviation (blue bars) for
$\mathcal{H}_1$ with $\phi_0=1/6$. The grey areas mark the frequency ranges
of the band gaps.
(d) $\bar{l}_e$ for $\mathcal{H}_3$ with $\phi_0=1/20$. In (a)-(d), the size of
the simulator $N=10$. The OAM included in the calculation is $l\in[-50,50]$.
Open and periodic
boundary conditions are used in the $x$ and $y$ direction. The photon loss
is $\gamma=0.1\kappa$ in (a) and (b), and $\gamma=0.2\kappa$ in (c) and (d).
}
\label{fig:edge-L}
\end{figure}

One of the most remarkable phenomena in topological physics is the existence of
topologically protected
chiral edge states in the band gaps of a finite lattice. In our system, we can
study the edge states by pumping the
cavity at the end of the \emph{1d} simulator array using a probing light beam
with a definitive OAM. It is equivalent to driving
a site on the edge of a $2d$ lattice. When the frequency of the probing light
falls in a band gap, excitation of gapless edge states dictates
that the light can only propagate along the edge of the simulated system. This
is clearly demonstrated in Figs. \ref{fig:edge-L} (a) and (b), where chiral
edge-state transport is observed in a small simulator.

To study the robustness of the edge states against disorder, we introduce the
average OAM ``displacement'' for the transport process defined as
\begin{eqnarray}
 \bar{l}_e(\omega)=\sum_{j_i\in edge}\sum_{j_o,l_o} \mathcal{T}_{j_i,0}^{j_o,
l_o}(\omega)\cdot l_o, \nonumber
\end{eqnarray}
where $\mathcal{T}_{j_i,0}^{j_o, l_o} =|T_{j_i,0}^{j_o,l_o}|^2$, and
$\sum_{j_i\in
edge}$ refers to summation over the sites close to one edge of the lattice where
the amplitude of the edge states is significant. As proved in the Supplementary
Note 5, when the frequency of the probing light $\omega$ falls in a large
band gap, $\bar{l}_e$ has the interesting property that it is
equal to the total Chern number for the bands below the gap. Also, the value of
$\bar{l}_e$ is mainly determined by states roughly in resonance with $\omega$.
Consequently, how $\bar{l}_e$ is disturbed by disorder is a measure for the
robustness of these states. Shown in Fig. \ref{fig:edge-L} (c) are $\bar{l}_e$
and its variation caused by a Gaussian distributed random shift $\delta\lambda$
in the cavity resonance frequency with a standard deviation
$\sigma(\delta\lambda)=0.1\kappa$. It can be concluded that the edge states are
almost immune to the disorder when the band gap is large compared to the photon
loss, whereas the in-band states are strongly affected.

In addition to its fundamental interest, edge-state transport is
also very useful for probing the topological behavior of a system. One
such example is the observation of the relativistic quantum Hall effect which
arises in the Dirac Hamiltonian $\mathcal{H}_3$ with
small but nonzero magnetic flux $\phi_0$. As shown in Fig.
\ref{fig:edge-L} (d), $\bar{l}_e$ experiences a double-step leap from $2$ to
$-2$ around the Dirac point at $\omega=0$ caused by a sudden change in the
topological property of the system. Such exotic behavior
\cite{bermudez10,mazza12} was predicted and
observed in graphene \cite{zhang2005experimental, goldman2009non}.

\subsubsection*{Topological quantum phase transition}
By measuring the system spectrum and edge-state transport, we can study
nontrivial physics such as topological quantum phase transitions driven by
non-Abelian gauge fields which are important
for understanding novel quantum
states of matter such as topological insulators and superconductors
\cite{hasan2010colloquium, NJP.12.033041,
NJP.14.015007, bermudez10, mazza12, PhysRevLett.105.255302, sheng2006quantum}.
In our system with non-Abelian gauge
field, if we choose $\hat\sigma_1= \sigma_x$, $\hat\sigma_2= \sigma_z$, $\phi_j
=0$,
$\alpha= 1/4$, $\beta_j= j/4+ \beta_0$, and $\lambda_j
=\lambda_0 \cdot[\text{mod}(j,4)-1.5]$ in equation (\ref{eq:H_param}), the
Hamiltonian in equation (\ref{eq:H_nonAbelian}) becomes
\begin{eqnarray}\label{eq:QSH}
\mathcal{H}_4=&-&\kappa\sum_{j,l}\left(\mathbf{\hat{a}}_{j,l+1}^{\dagger}e^{
i(\frac{\pi}{2}j+\beta_0)\sigma_z}\mathbf{\hat{a}}_{j,l}
+\mathbf{\hat{a}}_{j+1,l}^{\dagger}{i\sigma_x}\mathbf{\hat{a}}_{j
,l}+h.c. \right)
\nonumber \\
&+&\sum_{j,l}\lambda_0
\cdot[\text{mod}(j,4)-1.5]\mathbf{\hat{a}}_{j,l}^{\dagger}\mathbf{\hat{a}}_{j,l}
 \nonumber
\label{eq:H_TPT}
\end{eqnarray}
which describes an effective spin in a non-Abelian gauge field
characterized by spin-dependent magnetic field and strong spin-orbit coupling.
Also present is a periodically modulated on-site potential $\lambda_j$.
In the simulation system, the horizontal and vertical polarizations, which have
the same on-site energy, flip to their counter-part when the photon tunnels
between cavities and acquire opposite phases when the photon goes around a
plaquette in the simulated lattice in the same direction.
This is the same behavior with that of the spin up and down in an
electronic system which has time-reversal symmetry, and polarized photon edge
states analogous to spin edge states can emerge in our system. The two
polarized edge states are associated with opposite Chern numbers, and thus
their total Chern number
$C$ is 0 whereas the difference $\nu$ can be nonzero. The properties of such a
photonic topological insulator are in contrast with those of a normal insulator
in which both $C$ and $\nu$ are 0 and photon transport of both polarizations is
strongly suppressed.

\begin{figure}
\includegraphics[width=0.9\linewidth]{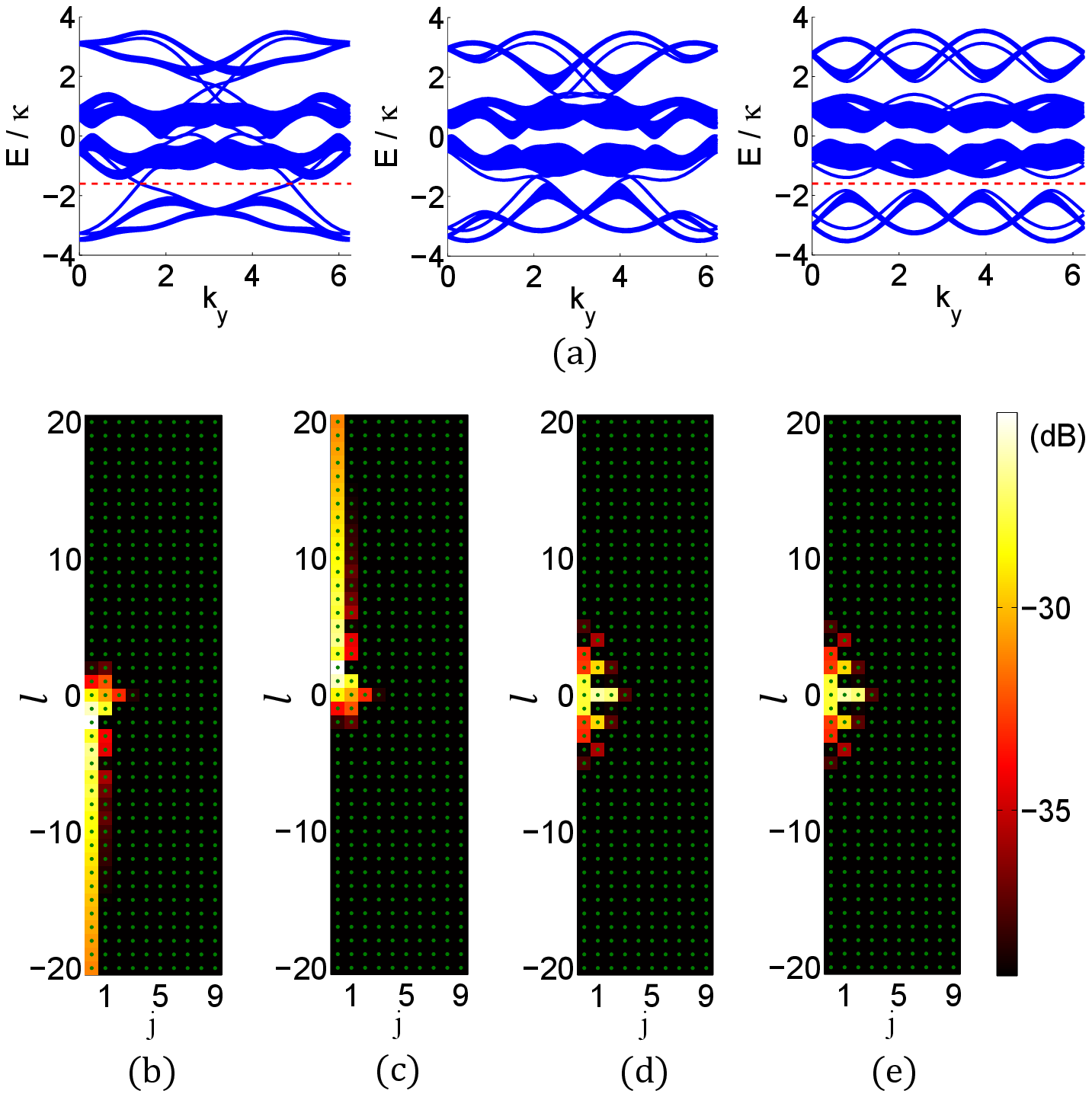}
\caption{Simulated topological quantum phase transition and edge-state
transport. (a) Calculated band structure for $\mathcal{H}_4$
with $\lambda_0=0.6\kappa$. The value of $\beta_0$ is $0$, $0.075$, and $0.125$
from left to right.
(b) Calculated photon transmission $\mathcal{T}_{0,0,s_i}^{j,l,s}$ when
$\beta_0=0$. The fist cavity in the simulator array is pumped by a probing light
with $l_i=0$ and $s_i=\leftrightarrow$. The frequency of the probing light
$\omega=-1.6\kappa$ is located in
the first band gap. (c) The same as in (b), except that the
polarization of the probing light is changed to the other value,
$s_i=\updownarrow$. (d) The same as in (b), except that $\beta_0 = 0.125$. (e)
The same as in (c), except that $\beta_0 = 0.125$.
In (b)-(e), the size of the simulator $N=10$. The OAM included in the
calculation is $l\in[-50,50]$. Open and periodic boundary
conditions are used in the $x$ and $y$ direction. The photon loss rate is
$\gamma=0.1\kappa$.}
\label{fig:TPT}
\end{figure}

A topological quantum phase transition can be induced in the
system by adjusting the value of the non-Abelian gauge field.
In Fig. \ref{fig:TPT} (a), it is shown how the band structure of the system
changes with $\beta_0$. As $\beta_0$ increases, the first band
gap near $\omega=-1.6\kappa$ closes and opens again. Initially, when $\beta_0$
is small, the topological
index $\nu$ of the system is $\nu=1$, and the system is in a topological
insulator state. Correspondingly, there are a pair of photon edge states with
opposite polarizations propagating in opposite directions as shown in Figs.
\ref{fig:TPT} (b) and (c). These polarized edge states are
protected as long as the local noise does not disturb the symmetry between the
two polarizations so that their on-site energies stay degenerate and
their phases around a plaquette remain opposite to each other.
When the energy gap opens again with a large $\beta_0$,
$\nu$ changes to 0, and the system becomes a normal insulator. This is
confirmed by the disappearance of the photon edge states in Figs.
\ref{fig:TPT}
(d) and (e).

\subsubsection*{Measurement of the Chern number}
\begin{figure}
\includegraphics[width=1.0\linewidth]{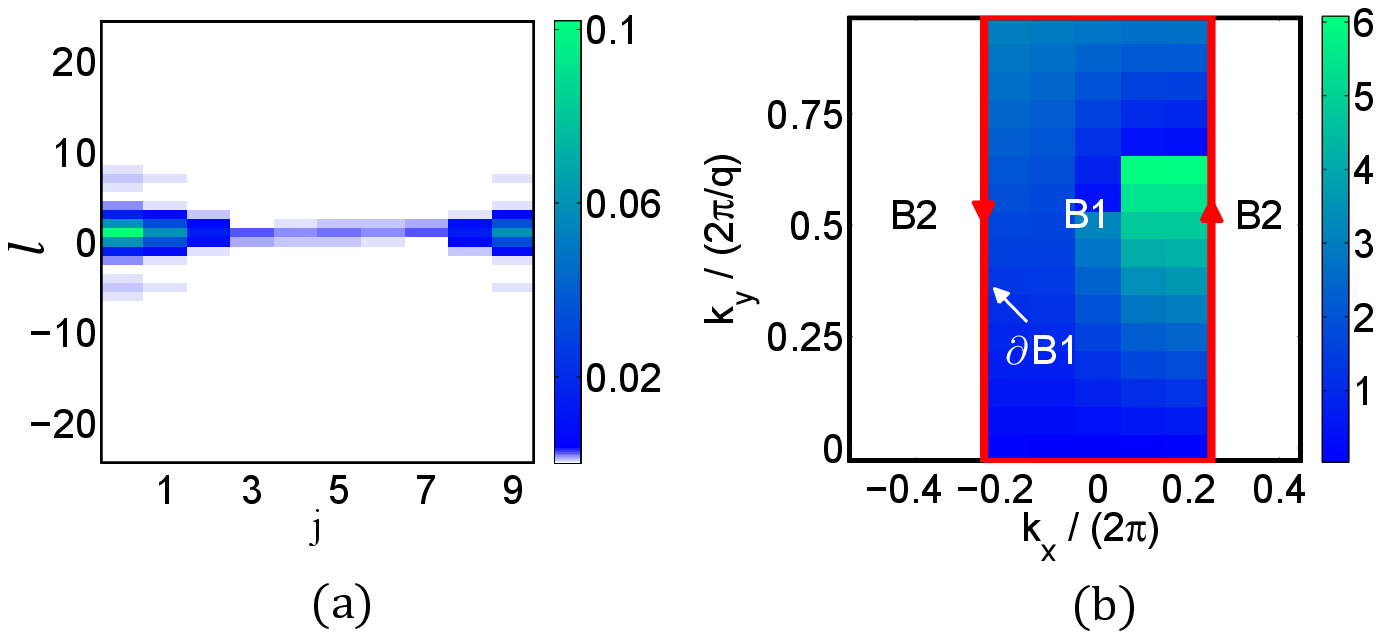}
\caption{Simulation of the Chern number measurement for
$\mathcal{H}_5$ with $\phi_0=1/6$.  (a) Calculated photon transmission
$|T_{0,0}^{j,l}|^2$. The probing light frequency is at
$\omega=-3.09\kappa$, where
lies the very narrow first energy band.
(b) Calculated phase mismatch $\chi(k_x,k_y)$ of $u_3^1$ in the
Brillouin zone resulting from two different phase conventions for
$u_l^1(k_x,k_y)$ in equation (\ref{eq:Bloch}), defined by dividing the Brillouin
zone into B1, where $u_0^1$
is always nonzero, and B2, which contains all zero points of $u_0^1$ but where
$u_3^1$ does not vanish. In one phase convention, $u_0^1$ is real in B1. In the
other convention, $u_3^1$ is real in B2. The Chern number is determined by the
integration of $\chi(k_x,k_y)$ on the boundary of B1, $\partial$B1
(Supplementary Note 6). In (a) and (b), the size of the
simulator $N=10$. The OAM included in the calculation is
$l\in[-48,48)$. Periodic boundary conditions are
used in both the $x$ and
$y$ directions. The photon loss rate $\gamma=0.1\kappa$.}
\label{fig:Chern}
\end{figure}

The Chern number is the ultimate quantum invariant to classify
topological states and characterize their behavior
\cite{hasan2010colloquium}. As shown in Fig.
\ref{fig:edge-L} (c), in a finite lattice the Chern number can be measured via
the average OAM displacement $\bar{l}_e$ for edge-state transport. In an
infinite system, the Chern number is equivalent to the TKNN index
\cite{thouless1982quantized}. For its measurement, we insert
a pair of beam rotators (BRs) with opposite rotation angles $\pm\vartheta =\pm
2\pi\phi_0$ in the coupling cavities, as shown in Fig.
\ref{fig:simulator_Abelian} (c). A BR with a rotation angle $\vartheta$
is made of two Dove prisms rotated by $\vartheta/2$ with respect to
each other and can change the azimuthal dependence of the OAM mode from
$e^{il\phi}$ to $e^{il(\phi+\vartheta)}$. We also
balance the two paths of the auxiliary cavities containing the SLMs.
The simulated Hamiltonian becomes
\begin{eqnarray}
\mathcal{H}_5=-\kappa\sum_{j,l}\left(a_{j,l+1}^{\dagger}a_{j,l}+
e^{-i2\pi l\phi_0}a_{j+1,l}^{\dagger}a_{j,l}+h.c. \right) \nonumber
\label{eq:Chern}
\end{eqnarray}
which is related to $\mathcal{H}_1$ by a gauge transformation and helps keep
the size of the simulator small (Supplementary Note 2). In Fig.
\ref{fig:Chern} (a), the amplitude of the photon transmission coefficients
$|T_{0,0}^{j,l}|^2$ is shown for a system with a rational magnetic flux
$\phi_0=1/6$. Similar to a situation described in
\cite{PhysRevLett.112.133902}, in a lossy cavity the probing light will be in
resonance with the entire first energy band of this system which is very narrow
(see Supplementary Note 6). This allows us to determine the in-band Bloch
eigenstates
\begin{equation}
e^{ik_xj}e^{ik_yl}u_l^1(k_x,k_y)
\label{eq:Bloch}
\end{equation}
from the Fourier transforms of $T_{0,0}^{j,l}$, where $k_x\in[-\pi,\pi]$,
$k_y\in[0,2\pi/6]$ define the Brillouin zone and
$u_{l}^m(k_x,k_y)=u_{l+6}^m(k_x,k_y)$ for the $m$-th band is a periodic
function. There is a Chern-number-conserving gauge freedom in the phase choices
of $u_l^1(k_x,k_y)$, as shown in Fig. \ref{fig:Chern} (b). $\chi(k_x,k_y)$, the
phase mismatch of $u_3^1$ resulting from the two different phase conventions in
Fig. \ref{fig:Chern} (b), can be used to calculate the Chern number
(Supplementary Note 6). Our numerical calculation using $\chi(k_x,k_y)$ yields
the Chern number 1 for the related band.

\section*{Discussion}
By mapping the OAM states of photons to spatial coordinates of a
lattice, we have found a promising scheme for studying nontrivial \emph{2d}
topological physics in a \emph{1d} physical simulator. Our method relies
on only linear optics and manipulation of OAM states, and thus it can be
realized with any physical systems that provide these elements or their
equivalent, though longer wavelengths may
have an advantage in coupling a large number of cavities. Our system is ready for
immediate experimental exploration, because the key
elements in our scheme, such as reliable manipulation of photon modes with
high angular momenta \cite{lee2004optical, fickler2012quantum}, precise
measurement of the OAM states \cite{malik2014direct, yao2011orbital}, design and
operation of degenerate cavities \cite{arnaud1969degenerate,
chalopin2010frequency}, and locking of multiple optical cavities \cite{xu12www},
have all been realized. Our idea may also be used to simulate 1d problems with OAM modes
in a single cavity \cite{PhysRevLett.102.135702, PhysRevLett.112.130401,
PhysRevLett.42.1698}, and it can lead to novel photonic effects with practical
applications \cite{PhysRevLett.100.013904}. Above all, by demonstrating the
counter-intuitive
application of photonic OAM in quantum simulation, our work deepens our
understanding of the OAM degree of freedom and advances our view of
photonic quantum simulation. Building upon the presented ideas, we can then
leverage the extreme flexibility and reliability in the
design and operation of optical circuits for quantum simulation of various
topological problems. All these issues and possibilities provide exciting
opportunities for further investigation.

\section*{Acknowledgements}
This work was funded by National Basic Research Program of China 2011CB921204,
2011CBA00200, the Strategic Priority Research Program of the Chinese Academy of
Sciences (Grant No. XDB01000000), National Natural Science Foundation of China
(Grant Nos. 11174270, 61490711, 11274289, 11325419, 61327901, 11274297, 61322506)
and the Fundamental Research Funds for the Central Universities (WK2470000011).
Z.-W.Z gratefully acknowledges the support of the K.C. Wong Education Foundation,
Hong Kong.



\bibliographystyle{unsrt}

\end{document}


\title{Supplementary Information for ``Quantum simulation of 2d topological
physics using orbital-angular-momentum-carrying photons in a 1d array of
cavities''}


\maketitle

\begin{figure}[!h]
\renewcommand\figurename{Supplementary Figure}
\includegraphics[width=0.35\linewidth]{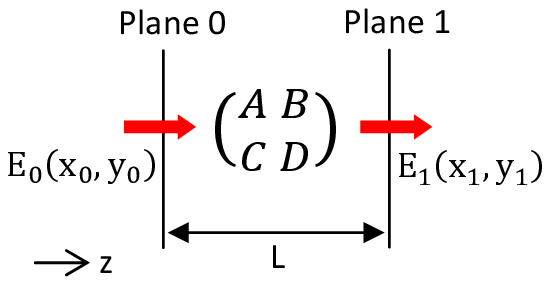}
\caption{Propagation of the light field between two planes perpendicular to the
optical axis in a cavity.}
\label{fig:collins}
\end{figure}

\begin{figure}[!h]
\renewcommand\figurename{Supplementary Figure}
\centering
\begin{minipage}[b]{1.0\textwidth}
\includegraphics[width=0.55\textwidth]{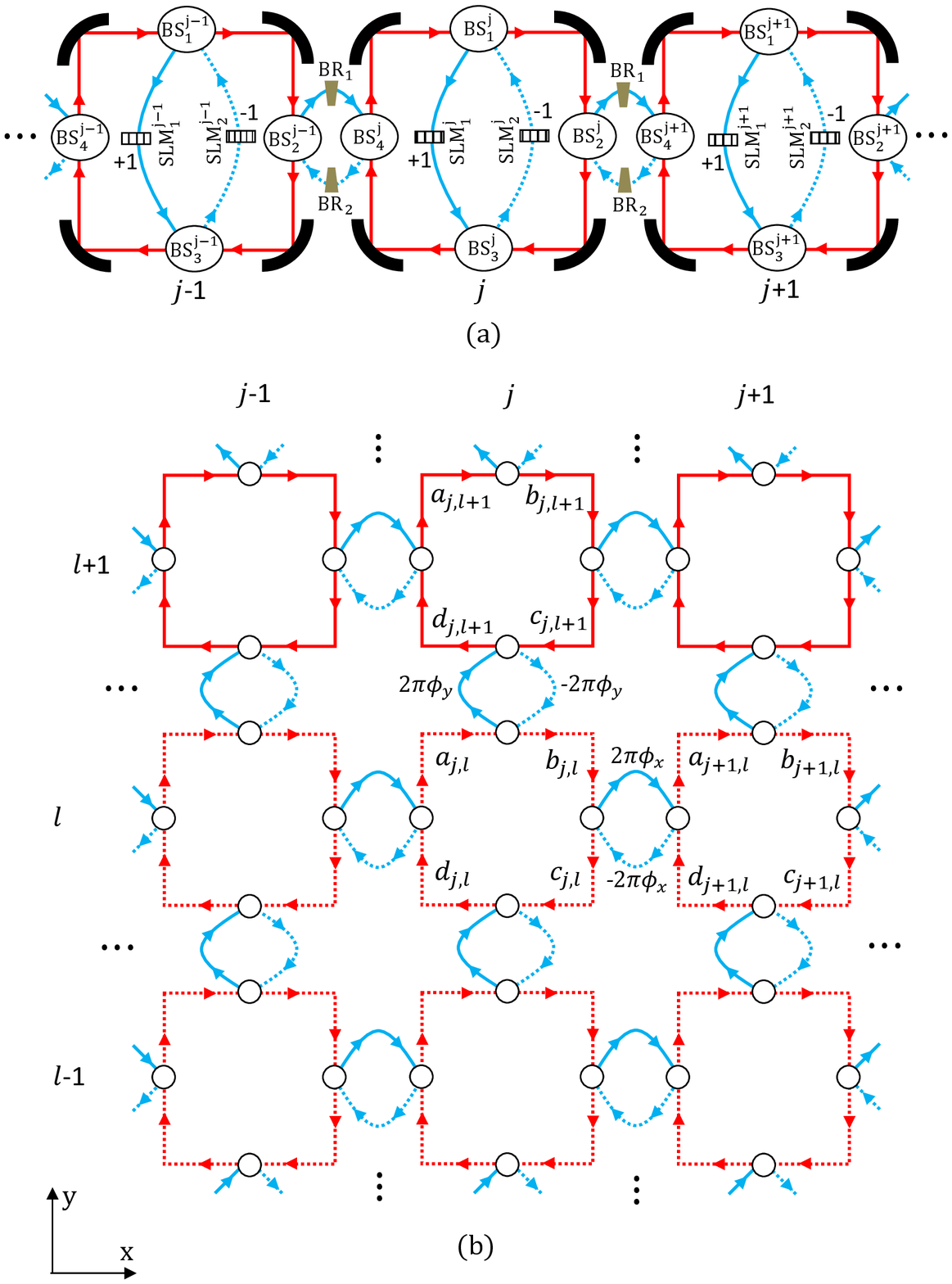}
\end{minipage}
\caption{(a) The simulator consisting of a \emph{1d} array of cavities.
Adjacent cavities are coupled by beam splitters (BSs). The SLMs in each cavity
change the OAM number of the photon by $\pm 1$. The beam rotators BR$_1$ and
BR$_2$, which have opposite rotation angles, can be use to implement
a gauge transformation of the magnetic field (see Supplementary Note 2). Their detailed design is shown in Supplementary Figure \ref{fig:BR}.
(b) The simulated \emph{2d} lattice system. For the convenience of discussion,
it is assumed that the BSs are placed at equal distances along the optical path
of the main cavity. The field amplitudes $a_{j,l}$,
$b_{j,l}$, $c_{j,l}$ and $d_{j,l}$ are defined at the mid point between adjacent
pairs of BSs. $2\pi\phi_x$ ($2\pi\phi_y$) is the phase imbalance between the
two arms of the corresponding cavity.}
\label{fig:system}
\end{figure}

\newpage

\begin{figure}[!h]
\renewcommand\figurename{Supplementary Figure}
\includegraphics[width=0.6\linewidth]{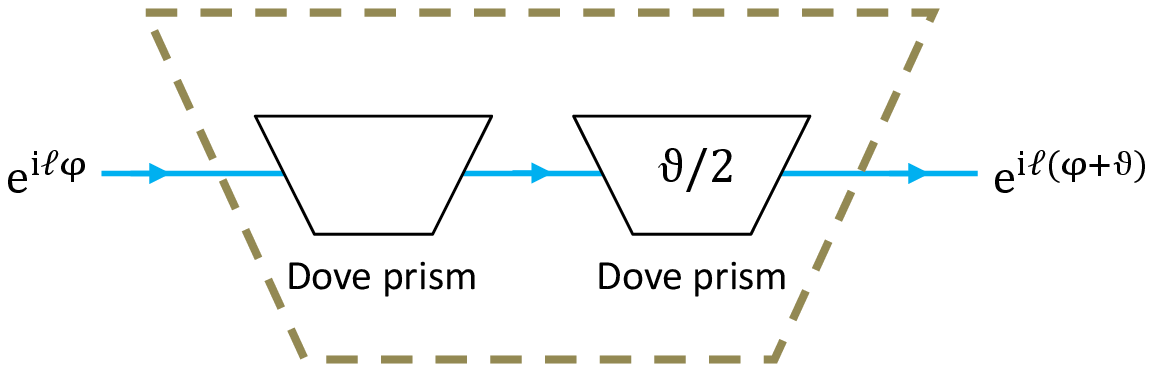}
\caption{A beam rotator consisting of two Dove prisms which are rotated by
$\vartheta/2$ with respect to each other. Since a Dove prism flips the
transverse profile of any transmitted beam, the two Dove prisms in the figure will rotate a
propagating beam by an angle $\vartheta$. It then changes the azimuthal phase
dependence of the $l$-th OAM mode from $e^{il\varphi}$ to
$e^{il(\varphi+\vartheta)} = e^{il\vartheta}e^{il\varphi}$.}
\label{fig:BR}
\end{figure}

\begin{figure}[!h]
\renewcommand\figurename{Supplementary Figure}
\includegraphics[width=0.8\linewidth]{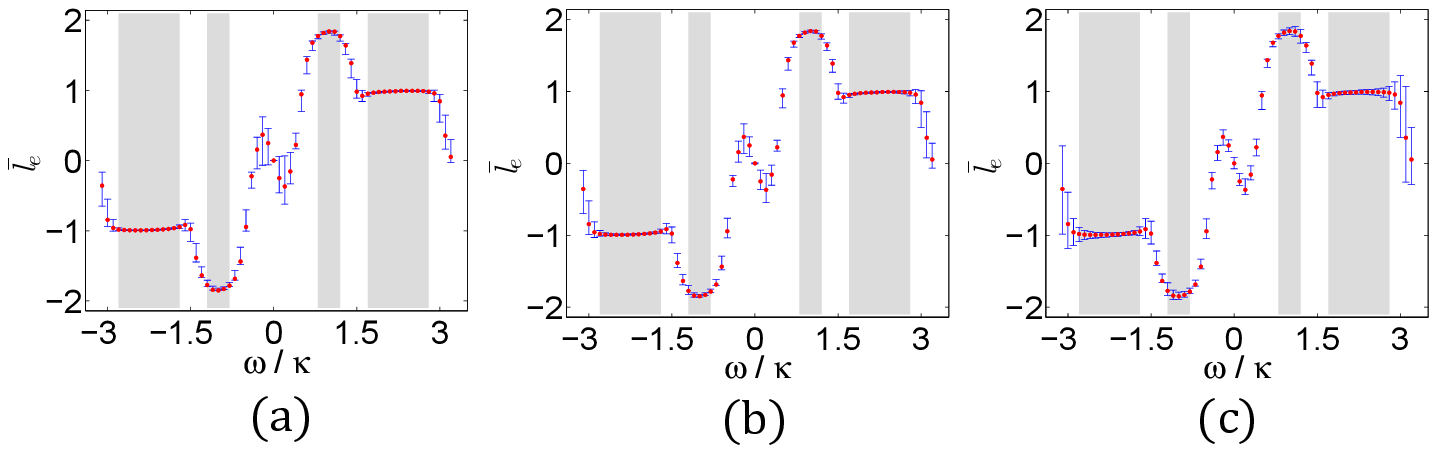}
\caption{The calculated average OAM displacement (red dots) for the photon
transmission and its standard deviation (blue bars) as a result of errors in
coupling strengths and photon loss for $\mathcal{H}_1$ with
$\phi_0=1/6$. The grey areas mark the frequency range of
the band gaps. (a) Uncertainties in both the magnitude and phase of $\kappa$ in
the $x$ direction are considered. They are assumed to have a Gaussian
distribution with a standard deviation of $\Delta |\kappa| = 0.05|\kappa|$ and
$\Delta\phi_{\kappa} = 0.05$rad. (b) OAM dependent uncertainties are
considered, by assuming an error of $\delta|\kappa|\cdot F(l+\frac{1}{2})$ in the
coupling between the $l$ and $l+1$ mode and an error of $\delta\gamma \cdot F(l)$ in
the photon loss for the $l$ mode, where $F(x)=1-e^{-\left(\frac{x}{30}\right)^2}$ and the
uncertainties have Gaussian distributions with standard deviations of $\Delta
|\kappa| = 0.05|\kappa|$ and $\Delta\gamma = 0.02\gamma$. (c) Independent uncertainties in couplings between OAM modes and
photon loss for each cavity are considered. The result is averaged over
input light with OAM number up to $\pm 3$. The distribution and standard
deviations of the uncertainties are the same as in (b). In (a)-(c),
the size of the simulator $N=10$. The OAM included in the calculation is $l\in
[-50,50]$. Open and periodic boundary conditions are used in the $x$ and $y$
direction. The photon loss rate is $\gamma = 0.2\kappa$.}
\label{fig:delta_kappa}
\end{figure}

\newpage

\begin{figure}[!h]
\renewcommand\figurename{Supplementary Figure}
\includegraphics[width=0.6\linewidth]{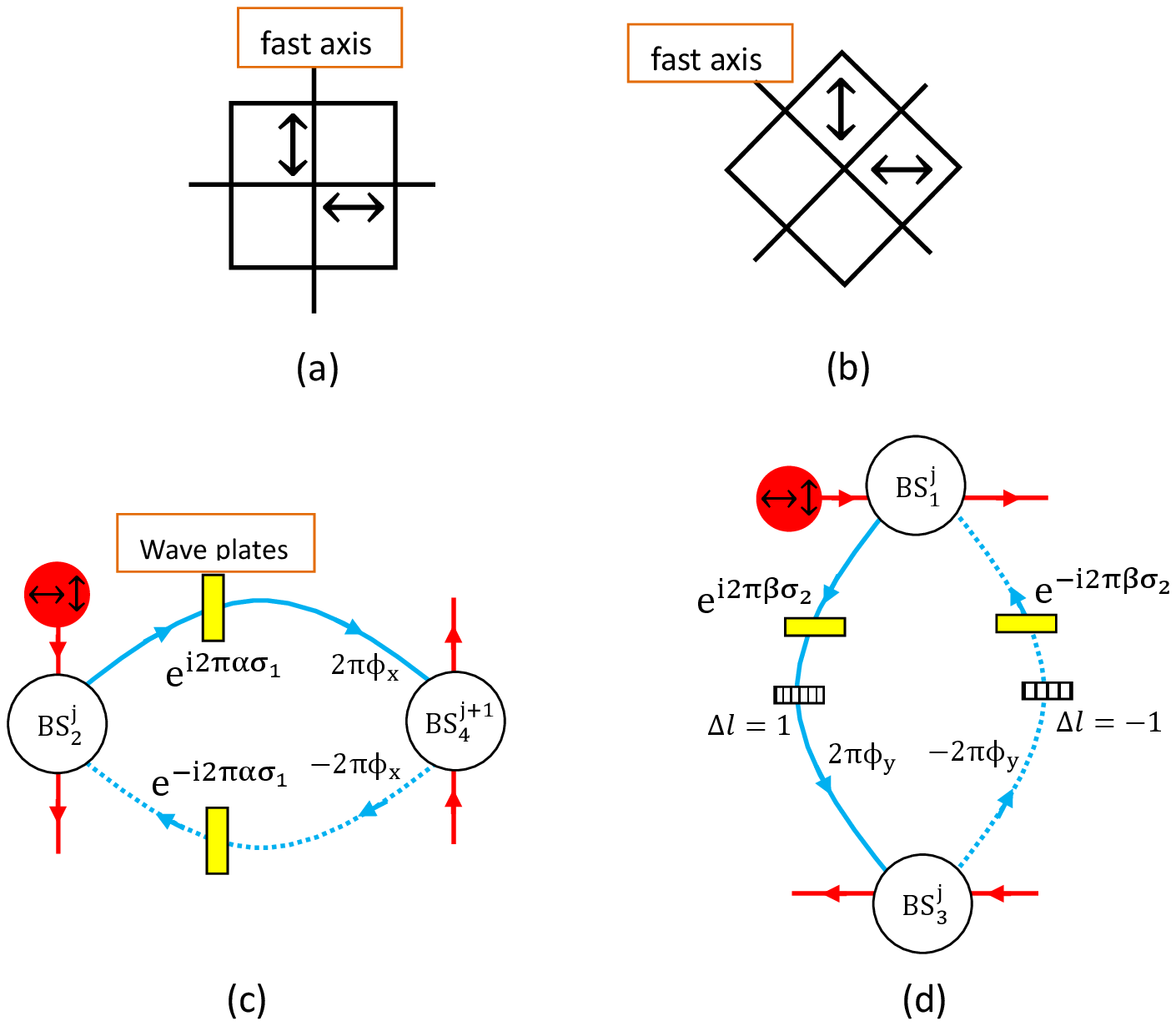}
\caption{(a) A waveplate whose fast axis aligns with the vertical polarization
of the incident light. The Jones matrix is $e^{i2\pi\phi\sigma_z}$ with the
phase $2\pi\phi$ is dependent on the thickness of the waveplate.
(b) A waveplate whose fast axis is rotated by $45^{\circ}$ with respect to the
vertical polarization of the incident light. The Jones matrix is
$e^{i2\pi\phi\sigma_x}$.
(c) The coupling cavity in the $x$ direction. The waveplates are designed to
realize Jones matrices $e^{\pm i2\pi\alpha\sigma_1}$, where $\sigma_1
= \vec{\sigma} \cdot\mathbf{n_1}$ with $\mathbf{n_1}$ an
arbitrary unit vector. $\pm 2\pi\phi_x$ is the spin-independent phase imbalance.
(d) The auxiliary cavity with the SLMs to change the OAM number of the photons.
The waveplates are designed to realize Jones matrices $e^{\pm
i2\pi\beta\sigma_2}$, where $\sigma_2
= \vec{\sigma} \cdot\mathbf{n_2}$ with $\mathbf{n_2}$ an
arbitrary unit vector. $\pm 2\pi\phi_y$ is the spin-independent
phase imbalance.}
\label{fig:non-abel-circuit}
\end{figure}

\begin{figure}[!h]
\renewcommand\figurename{Supplementary Figure}
\includegraphics[width=0.8\linewidth]{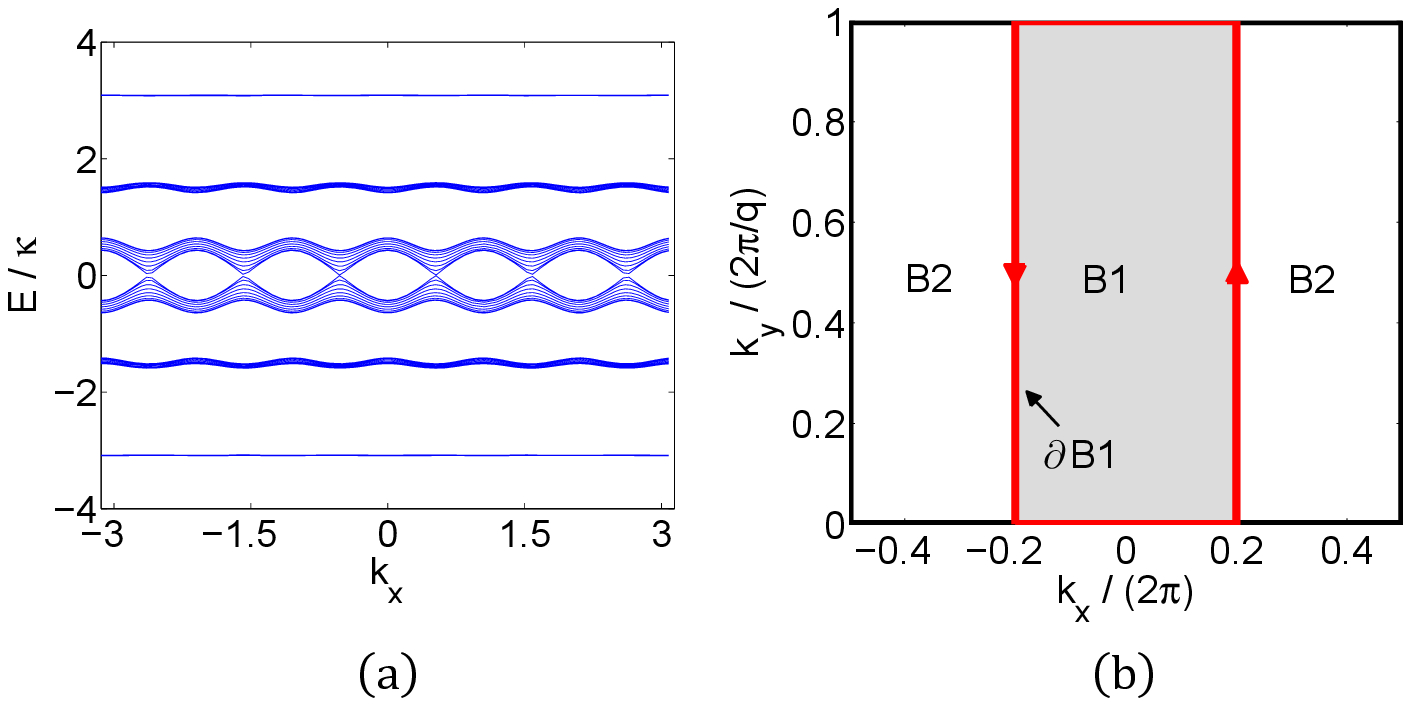}
\caption{
(a) The energy band structure of $\mathcal{H}_2$ in Eq. (\ref{eq:H_Abelian})
with a magnetic flux $\phi_0=1/6$. At each $k_x$, the eigenenergies for all
possible values of $k_y$ are calculated and plotted. It is seen that the bands
around $E=\mp 3.09\kappa$ are very narrow (they contain all eigenenergies and
are not a single line as they appear to be in the figure), with a width much
less than $0.1\kappa$.
(b) Division of the magnetic
Brillouin zone for the first band ($m=1$) around $E=-3.09\kappa$. B1 is the
area $\{k_x \in [-0.4\pi,0.4\pi], ky \in [0,2\pi/q]\}$. The rest is B2.
}
\label{fig:ChernS}
\end{figure}

\newpage

\section*{Supplementary Note 1: OAM modes in degenerate optical cavities}
All optical cavities in our simulation system are degenerate cavities that can
support optical modes with different orbital angular momentum (OAM). To
understand the design principles of
such cavities, we consider propagation of the light field in a cavity between
two planes perpendicular to the optical axis as depicted in Supplementary Figure
\ref{fig:collins}. For a cavity made of optical elements with rotational
symmetry, under the paraxial approximation, the position and slope of a ray at
the two planes, $[r_0, \dot{r}_0]^{T}$ and $[r_1, \dot{r}_1]^{T}$, are related
by \cite{hodgson2005laser}
\begin{equation}
\left[\begin{array}{c}
r_1\\
\dot{r}_1\end{array}\right]
=M\left[\begin{array}{c}
r_0\\
\dot{r}_0\end{array}\right] = \left[\begin{array}{cc}
A & B\\
C & D\end{array}\right]
\left[\begin{array}{c}
r_0\\
\dot{r}_0\end{array}\right],
\end{equation}
where the ray transfer matrix $M$ between the two planes is determined by the
optical design of the cavity. The electric fields at the two planes
are also related by the Collins integral \cite{collins1970lens}
\begin{equation}\label{eq:Collins}
e^{-ikz_1}E_1(x_1,y_1)=e^{-ikL}e^{-ikz_0}\frac{i}{\lambda B}\int\int
E_0(x_0,y_0)
\text{exp}[-\frac{i}{\lambda
B}(Ax_0^2+Dx_1^2-2x_0x_1+Ay_0^2+Dy_1^2-2y_0y_1)]dx_0dy_0,
\end{equation}
where $\lambda$ and $k$  are the wavelength and wave number, and $L$ is the
length of the optical path along the optical axis between the two planes.

The resonance frequencies and eigenmodes of the cavity can be solved for by
using the condition that the field must reproduce itself after a round trip in
the cavity. If the optical elements have cylindrical symmetry, the solutions
are the Laguerre-Gaussian (LG) modes $E_{p,l}(r,\varphi)e^{-ikz}$
\cite{hodgson2005laser} with the transverse field
\begin{eqnarray}
E_{p,l}(r,\varphi)=E_0\frac{W_{0}}{W(z)}\left(\frac{r\sqrt{2}}{W(z)}\right)^{|l|}
\mathcal{L}_p^ { |l| }\left(\frac{2r^2}{W(z)^2}\right) \nonumber\\
\times
\text{exp}\left(\frac{-r^2}{W(z)^2}\right)\text{exp}\left(\frac{-ikr^2}{2R(z)}
\right)\nonumber \\
\times \text{exp}\left[i(2p+|l|+1)\zeta(z)\right]e^{il\varphi},
\end{eqnarray}
where
$W(z)=W_0\sqrt{1+(z/z_0)^2}$ is the transverse width of the light beam,
$R(z)=z[1+(z_0/z)^2]$ is the wavefront curvature radius,
$\zeta(z)=\arctan(z/z_0)$ is the Gouy phase with beam waist $W_0$ and
Raleigh range $z_0=\pi W_0^2/\lambda$, and
$\mathcal{L}_p^{|l|}(x)$ is the generalized Laguerre polynomial.
The radial and azimuthal mode index $p$ and
$l$ determine the transverse distribution of the electric field, since $p + 1$
is the number of radial nodes and $2\pi l$ is the phase variation for a closed
path around the beam center. The resonance frequency for each $E_{p,l}$ mode in
a ring-type cavity is determined by \cite{arnaud1969degenerate}
\begin{equation}
kL_0 - (2p+l+1)\arccos \frac{A+D}{2} = 2n\pi,
\label{eq:mode_freq}
\end{equation}
where $n$ is an integer, $L_0$ is the length of the round-trip optical path,
and $A$ and $D$ are diagonal elements in the round-trip ray matrix. The
off-diagonal elements of the round-trip ray matrix, $B$ and $C$,
only affect the beam waist $W_0$ of the resonance modes.

It is seen from Eq. (\ref{eq:mode_freq}) that, generally speaking, different
$E_{p,l}$
modes are non-degenerate even for the same mode number $n$. However, If the
cavity is properly designed such that $A=D=1$ and $B=C=0$, the resonance frequency
becomes independent of the radial and azimuthal mode index $p$ and $l$. Such a
cavity is called a degenerate cavity. It can support photon modes of
different $p$ and $l$ simultaneously. The design requirement of degenerate
cavities is well understood; both general rules and concrete examples can be
found in the literature \cite{arnaud1969degenerate,
gigan2005image, chalopin2010frequency}.

Since each photon in a light beam with an azimuthal phase dependence
$e^{il\varphi}$ carries an OAM of $l\hbar$ \cite{allen1992orbital}, we can have
photons with different OAM in a degenerate cavity.
In our simulator shown in Supplementary Figure \ref{fig:system} (a),
there are three types of cavities with different
roles to form a \emph{1d} periodic array. Their optical design
is as follows.

\begin{enumerate}
 \item The main cavity in the array. Its length is chosen for constructive
interference, $kL_0=2n\pi$. The elements for the half round-trip ray matrix of
the optical
paths $BS_4^j\rightarrow BS_1^j\rightarrow BS_2^j$ and $BS_2^j\rightarrow
BS_3^j\rightarrow BS_4^j$ in Supplementary Figure \ref{fig:system} are
$A=D=-1$, $B=C=0$.

\item The coupling cavity between two adjacent main cavities consisting of
$BS_2^j$ and $BS_4^{j+1}$. Its length is chosen for destructive interference,
$kL_0=(2n+1)\pi$.
The elements of the ray matrix for the optical paths $BS_2^j\rightarrow
BS_4^{j+1}$ and $BS_4^{j+1}\rightarrow BS_2^{j}$ are $A=D=-1$, $B=C=0$.

 \item The auxiliary cavity consisting of the two beam splitters $BS_1^j$,
$BS_3^j$ and the two spatial light modulators $SLM_1^j$, $SLM_2^j$. Its length
is chosen for destructive interference, $kL_0=(2n+1)\pi$. The elements of the
ray matrix for optical paths $SLM_1^j\rightarrow BS_3^j\rightarrow SLM_2^j$ and
$SLM_2^j\rightarrow BS_1^j\rightarrow SLM_1^j$ are $A=D=-1$,
$B=C=0$.
\end{enumerate}

\section*{Supplementary Note 2: The tight-binding Hamiltonian}

\subsubsection*{Derivation of the Hamiltonian}
As explained in the main text, the \emph{1d} simulator in Supplementary Figure \ref{fig:system}
(a) is conceptually equivalent to the \emph{2d} rectangular lattice in Supplementary Figure
\ref{fig:system} (b). In order to derive the Hamiltonian of the simulated
system, we consider the eigenmode field $E$ which satisfies the Maxwell
equation
\begin{equation}\label{eq:TB1}
\nabla\times(\nabla\times E)=\epsilon(\mathbf{r})\frac{\omega^2}{c^2}E,
\end{equation}
where $\epsilon(\mathbf{r})$ is the dielectric constant of the system and
$\omega$ is the eigenenergy.

Under the assumption of weak coupling between cavities, $E$ can
be expanded in local modes (Wannier modes) \cite{poon2004designing,
bayindir2000tight, hartmann2006strongly},
\begin{equation}\label{eq:TB2}
E=\sum_{j,l}\psi_{j,l}W_{j,l}(\mathbf{r}),
\end{equation}
where $j$ is the index of the cavity in the simulator
array and $l$ is the OAM number of the photon.  $W_{j,l}$, the Wannier mode
localized at site $(j,l)$, satisfies the Maxwell equation
\begin{equation}\label{eq:TB3}
\nabla\times(\nabla\times W_{j,l})=\epsilon_0(\mathbf{r}-\mathbf{R}_{j,l})\frac{\omega_0^2}{c^2}W_{j,l}
\end{equation}
and is normalized to unity according to
\begin{equation}
\int d\mathbf{r}\epsilon_0(\mathbf{r}-\mathbf
{R}_{j,l}) W_{j,l}^*W_{j,l}=1
\end{equation}
with $\epsilon_0(\mathbf{r}-\mathbf{R}_{j,l})$ the dielectric
constant at site $(j,l)$, $\omega_0$ the single-site resonance frequency, and
$\mathbf{R}_{j,l}=j\hat \mathbf{x} + l \hat \mathbf{y}$ the lattice
vector at site $(j,l)$.

Using Eqs. (\ref{eq:TB1}), (\ref{eq:TB2}), and (\ref{eq:TB3}), we obtain
\begin{equation}\label{eq:TB4}
-\sum_{j',l'}\kappa_{j,l;j',l'}\psi_{j',l'}=(\omega-\omega_0)\psi_{j,l},
\end{equation}
where
\begin{equation}\label{eq:TB5}
\kappa_{j,l;j',l'}=\int
d\mathbf{r}\frac{\omega_0}{2}[\epsilon(\mathbf{r})-\epsilon_0(\mathbf{r}-\mathbf
{R}_{j',l'})] W_{j,l}^*W_{j',l'}.
\end{equation}
In deriving Eq. (\ref{eq:TB4}), we have used the weak
coupling condition $(\omega-\omega_0)/\omega_0, \kappa_{j,l;j',l'}/\omega_0 \ll
1$,
and kept only leading-order terms in $(\omega-\omega_0)/\omega_0$ and
$\kappa_{j,l;j',l'}/\omega_0$.
The on-site energy shift term $\kappa_{j,l;j,l}$ and non-adjacent coupling terms
are usually negligibly small compared to the coupling term between adjacent
cavities ($\kappa_{j,l;j+1,l}$ and $\kappa_{j,l;j,l+1}$), and we will drop
them.

In Eq. (\ref{eq:TB5}), the integration is limited to the region where
Wannier functions of neighboring cavities have appreciable overlap. In our
system, it is on the beam splitters that couple the cavities. Also, the
phase of the tunneling coefficient $\kappa_{j,l;j',l'}$ is sensitive to the
phase of the Wannier functions. We can see that, when there is a phase
imbalance $2\pi \phi_x$ between the two arms ($BS_2^j
\rightarrow BS_4^{j+1}$ and $BS_4^{j+1} \rightarrow BS_2^j$)
in the coupling cavity in Supplementary Figure \ref{fig:system}
(a), the phase shift of the Wannier function in the integration region with
respect to the balanced case $\phi_x=0$ results in the relation
\begin{equation}
\kappa_{j,l;j+1,l}(\phi_x)=\kappa_{j,l;j+1,l}(0) e^{i2\pi\phi_x},
\end{equation}
where $\kappa_{j,l;j+1,l}(0)$ is the tunneling coefficient for the balanced
case. Likewise, when the phase imbalance between the two paths (
$BS_1^j\rightarrow SLM_1^j
\rightarrow BS_3^j$ and $BS_3^j\rightarrow SLM_2^j \rightarrow BS_1^j$)
in the auxiliary cavities in Supplementary Figure \ref{fig:system} (a) is $2\pi \phi_y$, we have
\begin{equation}
\kappa_{j,l;j,l+1}(\phi_y)=\kappa_{j,l;j,l+1}(0) e^{i2\pi\phi_y},
\end{equation}
where $\kappa_{j,l;j,l+1}(0)$ is the tunneling coefficient in the $y$ direction
for the balanced case $\phi_y=0$. If we choose the same coupling strength in
the $x$ and $y$ direction, and denote $\kappa_{j,l;j+1,l}(0) = \kappa_{j,l;j,l+1}(0) = \kappa$,
Eq. (\ref{eq:TB4}) then leads to the
following tight-binding Hamiltonian in the rotating frame defined by
$\mathcal{H}_0=\sum \omega_0 \hat{a}_{j,l}^{\dagger}\hat{a}_{j,l}$,
\begin{eqnarray}\label{eq:tight-bind2}
\mathcal{H}=&-&\kappa\sum_{j,l}\left(e^{i
2\pi\phi_x}\hat{a}_{j+1,l}^{\dagger}\hat{a}_{j,l}
+e^{i 2\pi\phi_y}\hat{a}_{j,l+1}^{\dagger}\hat{a}_{j,l}+h.c.\right),
\end{eqnarray}
where $\hat{a}_{j,l}$ and $\hat{a}^\dagger_{j,l}$ are photon annihilation and
creation operators at site $(j,l)$. As discussed in the main text, if we choose
$\phi_x=0$, and $\phi_y$ to be linearly dependent on the index $j$ of the cavity
in the simulator array, $\phi_y =j\phi_0$, the corresponding Hamiltonian
\begin{eqnarray}
\mathcal{H}_1=&-&\kappa\sum_{j,l}\left(\hat{a}_{j+1,l}^{\dagger}\hat{a}_{j,l}
+e^{i 2\pi j\phi_0}\hat{a}_{j,l+1}^{\dagger}\hat{a}_{j,l}+h.c.\right)
\label{eq:H_B1}
\end{eqnarray}
describes a \emph{2d} system in a magnetic field with $\phi_0$ quanta of
flux per plaquette.

In some simulations we wish to introduce an on-site potential term to the
Hamiltonian. For this purpose, we can slightly detune the resonance frequency of
the main cavity from $\omega_0$. This results in the following additional term
in the Hamiltonian,
\[\sum_{j,l}\lambda_j\hat{a}_{j,l}^{\dagger}\hat{a}_{j,l},\]
where $\lambda_j=\omega_j-\omega_0$ and $\omega_j$ the resonance frequency of
the $j$-th main cavity.

\subsubsection*{Dependence of the tunneling coefficient on the BS reflectivity}
In order to select optical elements with appropriate parameters in experiments,
we need to understand how the tunneling coefficient $\kappa$ in Eq.
(\ref{eq:tight-bind2}) depends on the reflectivity of the BSs. This
can be accomplished by using the transfer matrix analysis
\cite{yariv2007photonics}. In Supplementary Figure \ref{fig:system} (b),
we introduce the photon field amplitudes $a_{j,l}$, $b_{j,l}$, $c_{j,l}$ and
$d_{j,l}$ at each lattice site $(j,l)$. We assume that the phase imbalances
$2\pi\phi_x$ and $2\pi\phi_y$ are the same for all lattice sites. In this case, the
system is periodic in both the $x$ and $y$ directions with a period of 1.
According to the transfer matrix formalism and Bloch theorem \cite{chremmos2008propagation,
chremmos2008modes},

\begin{equation}\label{eq:Bloch-condition}
\left(\begin{array}{c}
a_{j',l'}\\
b_{j',l'}\\
c_{j',l'}\\
d_{j',l'}\end{array}\right)=\left(\begin{array}{c}
a_{j,l}\\
b_{j,l}\\
c_{j,l}\\
d_{j,l}\end{array}\right)\cdot e^{-i(j'-j)K_x \Lambda-i(l'-l)K_y \Lambda},
\end{equation}
where $\Lambda$ is the unit spacing
 and $K_x$, $K_y$
are the Bloch quasi-momenta.

Assuming the reflection and transmission coefficients of all the BSs are $r=i|r|$
and $t=|t|$  ($|r|^2+|t|^2=1$), we can write their transfer matrix as
\begin{eqnarray}
M_{BS} & = & \left(\begin{array}{cccc}
\frac{1}{-i|r|} & \frac{t}{i|r|} \\
\frac{t}{-i|r|} & \frac{1}{i|r|} \\
\end{array}\right).\end{eqnarray}
Since the photons acquire a phase when they propagate between the BSs,
we have
\begin{equation}
\left(\begin{array}{c}
a_{j+1,l}\\
d_{j+1,l}\\
\end{array}\right)=M_x\cdot\left(\begin{array}{c}
b_{j,l}\\
c_{j,l}\\
\end{array}\right)
\end{equation}
with the field transfer matrix in the $x$ direction
\begin{equation}
M_x=\left(\begin{array}{cc}
e^{-ik S_c/8} & 0\\
0 & e^{ik S_c/8}\\
\end{array}\right)\cdot M_{BS}\cdot\left(\begin{array}{cc}
e^{-i(k S_a/2+2\pi \phi_x)} & 0\\
0 & e^{i(k S_a/2-2\pi\phi_x)}\\
\end{array}\right)\cdot M_{BS}\cdot\left(\begin{array}{cc}
e^{-ik S_c/8} & 0\\
0 & e^{ik S_c/8}\\
\end{array}\right),
\end{equation}
and similar expressions for $M_y$ in the $y$ direction. Here, $k$ is the wave
number, and $S_c$ and $S_a$ are the total optical path length of the main
cavity and the coupling cavity.
Using the Bloch relation in Eq. (\ref{eq:Bloch-condition}),
we can derive the following equations for the field amplitudes at site
$(j,l)$,
\begin{equation} \label{eq:bloch-mode1}
\left(\begin{array}{c}
a_{j,l}\\
d_{j,l}\\
\end{array}\right)=M_x\cdot\left(\begin{array}{c}
b_{j,l}\\
c_{j,l}\\
\end{array}\right)\cdot e^{iK_x \Lambda},
\end{equation}
and
\begin{equation}\label{eq:bloch-mode2}
\left(\begin{array}{c}
d_{j,l}\\
c_{j,l}\\
\end{array}\right)=M_y\cdot\left(\begin{array}{c}
a_{j,l}\\
b_{j,l}\\
\end{array}\right)\cdot e^{iK_y \Lambda}.
\end{equation}
By solving these equations, we obtain the Bloch modes and
dispersion relation of the system. The dispersion relation is given by
\cite{hafezi2011robust}
\begin{equation}\label{eq:dispersion}
\frac{\Omega_0}{\pi}\cdot\frac{|r|^2}{2}[\cos(K_x\Lambda- 2\pi\phi_x)+\cos(K_y\Lambda- 2\pi\phi_y)]=-(\omega-\omega_0) [1+O(|r|^2)],
\end{equation}
where $\Omega_0=2\pi \frac{c}{S_c}$ is the
free spectral range of the main cavity. Since the coupling is weak, $|r|^2\ll
1$, we can drop the higher order correction term $O(|r|^2)$.
Thus, from the dispersion relation in Eq. (\ref{eq:dispersion})
and the tight-binding Hamiltonian in Eq. (\ref{eq:tight-bind2}),
we get
\begin{equation}
\kappa=\frac{\Omega_0}{\pi}\cdot \frac{|r|^2}{4}.
\end{equation}

\subsubsection*{Gauge transformation}
It is well known that a magnetic field can be described by different vector
potentials which are related by a gauge transformation. This gauge
transformation can be implemented and tested in our system. As depicted in Supplementary Figure
\ref{fig:system} (a), we balance the lengths of the two optical paths in the
auxiliary cavities that contain the SLMs, and insert a pair of beam rotators
(BRs) with opposite rotation angles $\pm\vartheta =\pm 2\pi\phi_0$ in the two
arms
of the coupling cavities. The design of the BRs is shown in Supplementary Figure \ref{fig:BR}, where Dove prisms,
that flip the transverse profile of any transmitted beam \cite{born1980principles,
leach2002measuring}, are used.
By changing the azimuthal phase dependence of the $l$-th OAM mode from
$e^{il\varphi}$ to $e^{il(\varphi\pm 2\pi\phi_0)}$, they cause a phase shift
of $e^{\pm i2\pi l\phi_0}$ in the wave function when a photon tunnels between
two adjacent cavities. The simulated Hamiltonian then becomes
\begin{equation}
\mathcal{H}_2=-\kappa\sum_{j,l}\big(a_{j,l+1}^{\dagger}a_{j,l}+a_{j,l}^{\dagger}
a_{j,l+1}
+e^{-i2\pi l\phi_0}a_{j+1,l}^{\dagger}a_{j,l}+e^{i2\pi
l\phi_0}a_{j,l}^{\dagger}a_{j+1,l}\big),
 \label{eq:H_Abelian}
\end{equation}
which is a \emph{2d} system in a magnetic field with $\phi_0$ quanta of flux
per plaquette. $\mathcal{H}_2$ in Eq. (\ref{eq:H_Abelian}) is related to
$\mathcal{H}_1$ in Eq. (\ref{eq:H_B1}) by a gauge transformation.

Though $\mathcal{H}_2$ and $\mathcal{H}_1$ describe the same physics since they
are related by a gauge transformation, their implication for and requirement on
the simulation system can be quite different. When we are interested in bulk
properties (see Supplementary Note 6), a minimum number of unit cells in the
simulated 2d system are needed. Interestingly, this places different
requirements on the number of sites in both directions. It is because, for a
rational magnetic flux $\phi_0=p/q$ ($p$ and $q$ mutually prime integers), the
size of the magnetic unit cell is $1\times q$. Consequently, the system has a
period of 1 in one direction and $q$ in the other. Therefore, to simulate a
system with $M\times M$ magnetic unit cells, the size of the simulated system
should be $M\times qM$. Obviously, since the sizes in both directions are
different, we should choose a gauge in which the larger dimension is
represented with the degree of freedom that supports more sites. In our system,
the number of OAM modes in a cavity is much larger than the number of cavities
that can be coupled. This means that we should choose $\mathcal{H}_2$ to
minimize the size of the simulator (see Supplementary Note 6). It
requires $M$ cavities for simulating a system containing $M\times M$ magnetic
unit cells, whereas $qM$ cavities would have been needed if $\mathcal{H}_1$ was
chosen. As can be seen in this example, though $\mathcal{H}_2$ and
$\mathcal{H}_1$ are related by a gauge transformation and describe the same
physics, there is a major difference from the simulation point of view.

\subsubsection*{Characteristics of the simulated system in the $x$ and $y$
direction}

The characteristics of our simulated 2d systems are very different in the $x$
and $y$ direction because they are represented by completely different degrees
of freedom. In the $y$ direction, the sites of the lattice correspond to OAM
modes in the same cavity. Theoretically, since there is no upper limit for the
OAM of photons, the dimension in the $y$ direction is infinite. In practice,
properly designed degenerate cavities can accommodate many OAM modes, making the
number of sites in the $y$ direction very large. As can be seen from Supplementary Figure
\ref{fig:system} (a), neighboring OAM states in the same cavity are coupled by
the same set of BSs. Consequently, the coupling strengths between them are all
equal in theory. This is a huge advantage, and much better uniformity along the
$y$ direction can be achieved than what is possible in a chain of coupled
individual cavities whose sizes and separations will inevitably have
errors.

In the $x$ direction, multiple cavities need to be coupled in a chain. If
conventional optical cavities of macroscopic sizes are used at visible and
near-infrared wavelengths, the fluctuation in their lengths caused by thermal
noise and other disturbances can be comparable to the wavelength and it is
difficult to couple a large number of cavities. Nevertheless, because of the
importance of laser phase and frequency stabilization in many contexts, there has
been a long history of development of experimental techniques to deal with this
problem \cite{drever83}. By using advanced experimental techniques, it is now
possible to lock multiple cavities and perform sophisticated experiments
\cite{su12,yokoyama13}. As shown in the main text, to observe and study
topological effects in our system, we only need a small 1d array with just a few
cavities which is within the capability of current technologies. To increase
the number of cavities that can be coupled, one can use technologies with more
stable cavities, or work with photons with longer wavelengths such as microwave
or maser photons \cite{tamburini11,tamburini12}.

Another issue in the $x$ direction is with the coupling strength between
cavities. Since all OAM modes in the same cavity are eigen solutions of the
same wave equation, once 1 OAM mode in a cavity is locked with the
corresponding mode in the neighboring cavity, all other OAM modes are locked
too. Therefore, locking cavities with multiple OAM modes is not more difficult
than locking cavities with a single mode only. Still, coupling strengths
between different cavities can fluctuate since they are realized with different
optical elements.
Such fluctuations in the coupling strength between cavities have an adverse
impact on propagation of light through the body of the simulated lattice by
in-band bulk states, but they obviously do not disturb the edge-state transport
which is confined to the edge of the system. This is true as long as these
fluctuations are much smaller than the band gap of the system and do not destroy
its topology, a requirement not difficult to meet because of the availability
of BSs with very accurate reflectivities.
To see quantitatively how the simulation is
affected by errors in the coupling strength, we plot the average OAM
displacement (which is defined in equation (\ref{eq:l_e}) and shown to be
determined by the Chern number of the system) for the photon transmission
and its fluctuation caused by such errors in Supplementary Figure
\ref{fig:delta_kappa} (a). It can
be seen that edge-state transport in the band gaps is hardly disturbed by small
errors in the coupling strength between cavities.

\subsubsection*{OAM-Dependence of the tunneling coefficient and photon loss}

As mentioned above, the couplings between different OAM states in the same
cavity are realized with the same set of BSs and thus in principle they should
all be equal. This argument is complicated by the practical consideration that,
in reality, the SLMs have only limited resolution, and couplings between OAM
modes can be dependent on the OAM number $l$ because their spatial extends
are different, especially for high OAM modes. This is only an issue when the
photon loss is very low (otherwise very little light propagates to high OAM
modes). It can be dealt with by using high-resolution SLMs for which such
dependence is very weak. There are also
experimental techniques to minimize and eliminate such dependence. For instance,
it is experimentally demonstrated in \cite{oemrawsingh2004production} that the
spatial extends of the OAM modes can be made the same on two SLMs in the
optical path provided that appropriate optical design is used between them to
place them in each other's near fields. Similar techniques can be used in our
system to design the round-trip ray matrix such that the spatial extends of
the OAM modes return to their original value when they come back to the SLM
after a round-trip in the cavity following an increment/decrement in their OAM
number by the SLM.

Nevertheless, considering the many inevitable and uncontrollable uncertainties
in an actual experiment, the couplings between high OAM modes will likely have
some, albeit weak dependence on the OAM number despite the precautions taken.
The quality factors of the high OAM modes can depend on the mode number too,
since modes with different spatial extends will have different leakage. Due to
this OAM dependence, the characteristics of the component related uncertainties
in our system are different than those in a \emph{2d} cavity array where they
are independent for each cavity. Assuming the same magnitude for the
uncertainties in each case (though in reality the uncertainties in a \emph{2d}
cavity array are likely much greater when the size of the array is large), this
distinction in their characteristics should be insignificant, because
topological protection ensures that edge-state transport is not disturbed by
the uncertainties as long as they are much smaller than the band gap of the
system and thus do not destroy its topology. Though the exact dependence on the
OAM number is difficult to calculate, in a numeric simulation to check the
robustness of the edge-state transport we can assume any dependence
since topological protection is not sensitive to the exact form of the local
noise. In Supplementary Figure \ref{fig:delta_kappa} (b), we show the calculated average OAM
displacement for an ideal system without uncertainties and its fluctuations
caused by errors in the coupling strength and Q factors, assuming a particular
dependence on the OAM number which results in larger errors for higher OAM
modes.
As we can see, within the band gaps where the transport is via edge states, the
average OAM displacement is hardly disturbed by the OAM dependent errors. In
contrast, the in-band bulk state transport is strongly
affected. For comparison, we perform the same calculation for a \emph{2d}
cavity array and plot the results in Supplementary Figure \ref{fig:delta_kappa} (c), by assuming
the same magnitude of errors in the parameters though they are independent for
each cavity. As far as edge-state transport is concerned, there is no
appreciable difference between the two cases. Therefore, though in reality the
component related uncertainties in a large 2d cavity array are likely to be much
greater than in our system, under the assumption of similar magnitude for the
uncertainties the behavior of edge-state transport is the same.

\section*{Supplementary Note 3: Simulation system for non-Abelian gauge fields}

In order to simulate topological physics associated with non-Abelian gauge
fields, we use polarized photons and represent the spin up and down states with
the horizontal ($|\leftrightarrow\rangle$) and vertical
($|\updownarrow\rangle$) polarization. The photon modes are
$\mathbf{\hat{a}}^{\dagger}_{j,l}=(\hat{a}^{\dagger}_{j,l,\leftrightarrow},
\hat{ a}^{\dagger}_{j,l,\updownarrow})$, where
$\hat{a}^{\dagger}_{j,l,\leftrightarrow}$ and
$\hat{a}^{\dagger}_{j,l,\updownarrow}$ are the creation operators for
horizontally and vertically polarized photons at site $(j,l)$.

The design of the main cavities of the simulator does not require any
modification. The auxiliary and coupling cavities, however, need to be
augmented with polarization manipulating elements. Shown in
Supplementary Figure \ref{fig:non-abel-circuit} (a) and (b) are birefringent waveplates used in
the auxiliary and coupling cavities. Such wave plates can alter the
polarization state of the photons because polarization components along the fast
and slow axis travel at different speeds \cite{yariv2007photonics}. In Supplementary Figure
\ref{fig:non-abel-circuit} (a), when the fast axis of the waveplate aligns with
the vertical polarization of the incident photons, the two polarization states
acquire different phases after the photons pass through the waveplate
\cite{yariv2007photonics},

\begin{equation}
\left(\begin{array}{cc}
|\leftrightarrow\rangle\\
|\updownarrow\rangle
\end{array}\right)\Rightarrow e^{i2\pi\phi \sigma_z}
\left(\begin{array}{cc}
|\leftrightarrow\rangle\\
|\updownarrow\rangle\\
\end{array}\right),
\end{equation}
where $\sigma_z$ is the Pauli matrix, and $e^{i2\pi\phi\sigma_z}$ is the
corresponding Jones matrix with the phase $\phi$ dependent on the thickness of
the waveplate. If the fast axis is rotated by $45^{\circ}$ with respect to the
vertical polarization of the incident photons as in Supplementary Figure
\ref{fig:non-abel-circuit} (b), the corresponding Jones matrix becomes
$e^{i2\pi\phi\sigma_x}$.
Likewise, by taking advantage of the fact that left and right-handed circularly
polarized light travels at different speed in optical media with circular
birefringence, we can design a polarization rotator which has a
Jones matrix $e^{i2\pi\phi\sigma_y}$ \cite{yariv2007photonics}.
More generally, with a proper combination of waveplates and (or) rotators,
we can realize any desired Jones matrix $e^{i2\pi\phi\sigma_\mathbf{n}}$
\cite{yariv2007photonics}, where $\sigma_n = \vec{\sigma} \cdot\mathbf{n}$
and $\mathbf{n}=(n_x,n_y,n_z)$ is an arbitrary unit vector.

In Supplementary Figure \ref{fig:non-abel-circuit} (c), we design the coupling
cavities in the $x$ direction such that the optical paths $BS_2^j \rightarrow
BS_4^{j+1}$ and $BS_4^{j+1} \rightarrow BS_2^{j}$ contain phases
$\frac{kS_a}{2}\pm
2\pi\phi_x$ and Jones matrices $e^{\pm i2\pi\alpha\sigma_1}$,
where $\sigma_1 = \vec{\sigma} \cdot\mathbf{n_1}$ with $\mathbf{n_1}$ an
arbitrary unit vector. The Hamiltonian of the coupling term in the $x$
direction then reads
\begin{eqnarray}
-\kappa\sum_{j,l}\left(\mathbf{\hat{a}}_{j+1,l}^{\dagger}e^{i2\pi(\phi_{x}
+\alpha\sigma_1)}\mathbf{\hat{a}}_{j,l}+h.c.\right).
\label{eq:Hx_nonAbel}
\end{eqnarray}
The physical meaning of the phases is easier to understand if we switch to the
eigen polarization states of $\sigma_1$, $|\leftrightarrow'\rangle$
and $|\updownarrow'\rangle$. In these bases, Eq. (\ref{eq:Hx_nonAbel}) is
\begin{eqnarray}
-\kappa\sum_{j,l} \left(e^{i2\pi(\phi_{x}-\alpha)}
\hat{a}_{j+1,l,\leftrightarrow'}^{\dagger}\hat{a}_{j,l,\leftrightarrow'}+e^{
i2\pi(\phi_{x}+\alpha)}\hat{a}_{j+1,l,\updownarrow'}^{\dagger}\hat{a}_{j,l,
\updownarrow'}+h.c.\right).
\end{eqnarray}
Obviously, $2\pi(\phi_x\pm\alpha)$ are the tunneling phases for photons in
states $|\leftrightarrow'\rangle$
and $|\updownarrow'\rangle$ respectively.

The design of the polarization manipulating circuits for the auxiliary cavities
is shown in Supplementary Figure \ref{fig:non-abel-circuit} (d). The tight-binding Hamiltonian for the system is
then
\begin{eqnarray}
\mathcal{H}=&-&\kappa\sum_{j,l}\left(\mathbf{\hat{a}}_{j+1,l}^{
\dagger } e^ { i\hat {2\pi\theta}_x}\mathbf{\hat{a}}_{j,l}
+\mathbf{\hat{a}}_{j,l+1}^{\dagger}e^{i\hat{2\pi\theta}_y}\mathbf{\hat{a}}_{j,l}+h.c.\right) \nonumber \\
&+&\sum_{j,l}\lambda_j\mathbf{\hat{a}}_{j,l}^{\dagger}\mathbf{\hat{a}}_{j,l},
\label{eq:H_nonAbelian}
\end{eqnarray}
where $\lambda_j$ is the detuning of the $j$-th cavity, and the tunneling phases
are
\begin{eqnarray}
\hat{\theta}_x=\phi_x+\alpha{\sigma}_1, \hat{\theta}_y=\phi_y+\beta{\sigma}_2,
\label{eq:Peierls}
\end{eqnarray}
with $\sigma_2=\vec{\sigma} \cdot\mathbf{n_2}$ and $\mathbf{n_2}$ a unit vector.
$2\pi\phi_x$, $2\pi\phi_y$
are the spin-independent part of the gauge fields.

The spin-dependent $\hat{\theta}_x$ and $\hat{\theta}_y$ in Eq.
(\ref{eq:Peierls}) do not
necessarily commute \cite{dalibard2011colloquium}. When $\hat{\theta}_x\hat{\theta}_y\neq
\hat{\theta}_y\hat{\theta}_x$, they correspond to non-Abelian gauge potentials,
and the Hamiltonian in Eq. (\ref{eq:H_nonAbelian}) can be used to simulate the
effects of non-Abelian gauge fields.

Notice that the horizontal and vertical polarizations of light used in
our simulation system are both clockwise circulating cavity modes. By
assuming that there is no coupling between the clockwise and counterclockwise cavity modes,
and restricting
ourselves to clockwise cavity modes only, we can describe the behavior of the
horizontal and vertical polarizations with the non-Abelian Hamiltonian in
equation (\ref{eq:H_nonAbelian}). Since the Jones matrix description applies
to polarizations of light traveling in one direction, and we make use of
clockwise cavity modes only, we are not simulating the physical time-reversal
symmetry directly. Nevertheless, due to the optical setup
of the system, the phases acquired by and transitions between vertical and
horizontal polarizations are the same with those of spin up and down in an
electronic system described by the Hamiltonian in equation
(\ref{eq:H_nonAbelian}). Because of this, we can have polarized photon edge
states in our system which are topologically protected by the symmetry in the
optical design for the two polarizations though they are not physical
time-reversal conjugates.

\section*{Supplementary Note 4: Input-output formalism for photon transmission measurement}
As described in the main text of the paper, we probe our system by
coupling a light beam with a definitive OAM number (and polarization in
studies associated with non-Abelian gauge fields) to a cavity in the simulator
array and measuring the photon transmission to other OAM modes (and
polarizations when relevant) in different cavities. To study the characteristics
of the measured quantity, we now consider the transmission coefficient taking
into account the effect of photon loss.
The photon loss can be understood in terms of the coupling of the cavity modes
with the outside world due to a coupling term $\mathcal{H}_{INT}$ in the
system's total Hamiltonian
\begin{equation}
\mathcal{H}=\mathcal{H}_{SYS}+\mathcal{H}_{BATH}+\mathcal{H}_{INT},
\end{equation}
where $\mathcal{H}_{SYS}$, $\mathcal{H}_{BATH}$ are the Hamiltonian for the
cavity field and outside bath field. In the rotated frame with respect to
the resonance frequency of the cavity, we have
\begin{equation}
\mathcal{H}_{BATH}=\sum_n\int_{-\infty}^{+\infty}d\omega[\omega d^{\dagger}_n(\omega)d_n(\omega)],
\end{equation}
and
\begin{equation}
\mathcal{H}_{INT}=-i\sum_{n}\int_{-\infty}^{+\infty}d\omega
\sqrt{\frac{\gamma_n}{2\pi}}[d_{n}(\omega)a^{\dagger}_{n}-a_{n}d_{n}^{\dagger}
(\omega)].
\end{equation}
Here, $n=[j,l,s]$ is a collection of
quantum numbers to specify a cavity photon mode. It includes the index of the
cavity in the simulator array ($j$), the OAM number ($l$) of the photon, and its
polarization state
($s=\leftrightarrow,\updownarrow$) when relevant.
$\omega$ denotes the frequency detuning from the resonance
frequency $\omega_0$.
$d_{n}(\omega)$ is the
operator for the environment field coupled to the cavity photon mode labeled by
$n$. $d_{n}(\omega)$ obeys the commutation
relation
\[[d_{n}(\omega),d_{n'}^{\dagger}(\omega')]=\delta_{nn'}\delta(\omega-\omega').\]
The system Hamiltonian has a bilinear form
\begin{eqnarray}
\mathcal{H}_{SYS}&=&\sum_{n,n'} a^{\dagger}_nH_{nn'}a_{n'},
\end{eqnarray}
where $H_{nn'}$ is the matrix element of the simulated Hamiltonian $\mathcal{H}_{SYS}$.

Using the input-output theorem \cite{walls2008quantum}, we can write the
Langevin equation of the system operators,
\begin{eqnarray}
\frac{da_n(t)}{dt}&=&-i[a_n,\mathcal{H}_{SYS}]-\frac{\gamma_n}{2} a_n(t)-\sqrt{\gamma_n}d_{in,n}(t) \nonumber \\
&=& -i\sum_{n'}H_{nn'}a_{n'}(t)-\frac{\gamma_n}{2} a_n(t)-\sqrt{\gamma_n}d_{in,n}(t),
\end{eqnarray}
where $d_{in,n}(t)=\frac{1}{\sqrt{2\pi}}\int_{-\infty}^{+\infty}d\omega
e^{-i\omega t}d_{n,0}(\omega)$ is the input field operator, with
$d_{n,0}(\omega)$ the value of $d_{n}(\omega)$ at $t=0$.
The output field is obtained from the input-output
formalism
\[d_{out,n}(t)-d_{in,n}(t)=\sqrt{\gamma_n}a_n(t).\]
Making a Fourier transformation, we get
\begin{eqnarray}
-i\omega a_n(\omega)&=& -i\sum_{n'}H_{nn'}a_{n'}(\omega)-\frac{\gamma_n}{2}
a_n(\omega)-\sqrt{\gamma_n}d_{in,n}(\omega),\nonumber \\
\sqrt{\gamma_n}a_n(\omega)&=&d_{out,n}(\omega)-d_{in,n}(\omega).
\end{eqnarray}
The solution is
\begin{equation}
d_{out,n'}(\omega)=\sum_n\left\{\delta_{n'n}-i\left[\sqrt{\Gamma}\frac{1}{\omega-\mathcal{H}_{SYS}+i\Gamma/2}
\sqrt{\Gamma}\right]_{n'n}\right\} d_{in,n}(\omega),
\end{equation}
where $\Gamma=\text{diag}\{\gamma_1,\gamma_2,\gamma_3,\ldots\}$ is the decay matrix.
The first term on the right hand side, $d_{in,n'}(\omega)$, is the reflection.
The rest describes field transmission.
The transmission coefficient is
\begin{equation}
T_{n}^{n'}=-i\left[\sqrt{\Gamma}\frac{1}{\omega-\mathcal{H}_{SYS}+i\Gamma/2}\sqrt{\Gamma}\right]_{
n'n}.
\end{equation}
For the simple case when all cavity modes decay with the same rate
$\gamma_n=\gamma$ $(\forall n)$, the transmission coefficient is
\begin{equation}
T_{n}^{n'}=-i\langle
n'|\frac{\gamma}{\omega-\mathcal{H}_{SYS}+i\gamma/2}|n\rangle,
\label{eq:T_n}
\end{equation}
where $|n\rangle=\hat{a}^{\dagger}_n|0\rangle$ is a single photon state.

\section*{Supplementary Note 5: OAM displacement in edge-state transport}
It is demonstrated in the main text that, when the frequency of a probing
light falls in a gap in the spectrum of a finite \emph{2d} lattice in magnetic
field with the Hamiltonian
\begin{equation}
 \mathcal{H}=-\kappa\sum_{j,l}\left(e^{i2\pi j\phi_0}
 \hat{a}_{j,l+1}^{\dagger}\hat{ a}_{j,l}+\hat{a}_{j+1,l}^{\dagger}\hat{
 a}_{j,l},
 +h.c.\right),
 \end{equation}
it can only propagate along the edge of the lattice
because of edge-state excitation. We discovered a quantity that is very useful
for the study of edge-state transport.
It is the average OAM displacement defined as
\begin{equation}
 \bar{l}_e=\sum_{j\in \text{edge}}\sum_{j_o,l_o}|T_{j,0}^{j_o, l_o}|^2\cdot
l_o,
\label{eq:l_e}
\end{equation}
where $T_{j,0}^{j_o, l_o}$ is the photon transmission coefficient defined in
Eq. (\ref{eq:T_n}) and $\sum_{j\in \text{edge}}$ refers to summation over the
region close to one edge (left or right) of the lattice where the
amplitude of the corresponding edge states is appreciable.

It can be shown that $\bar{l}_e$ defined in Eq. (\ref{eq:l_e}) is related to
the Chern number of the system. To prove this, we consider a system in the
Laughlin-Halperin geometry which has open and periodic boundary condition in the
$x$ and $y$ direction. In such a system, there are two sets of chiral edge
states, one per boundary, that propagate in opposite
directions \cite{hatsugai1993edge, hatsugai1993chern}. Consequently, the
displacement $\bar{l}_e$ due to transport by edge states on the left and right
edges are equal in magnitude but opposite in sign. Without loss of
generality, we will focus on the left edge, and restrict the
summation of $j$ to the region near the left edge of the lattice.
Because of the periodic boundary condition in the $y$
direction, the Bloch momentum $k_y=2\pi\frac{n_y}{N_y}$ is a good quantum
number of the system, where $n_y=0,1,\ldots, N_y-1$ and $N_y$ is the number of
sites in the $y$ direction. We can use the momentum representation in the
$y$ direction,
$\hat{a}^{\dag}_{j,k_y}=\frac{1}{\sqrt{N_y}}\sum_{l}e^{ik_yl}\hat{a}^
{\dag}_{j, l}$, and introduce the single-particle eigenfunction
\begin{equation}
|\Psi_{k_y}\rangle=\sum_j \Psi_{j,k_y}\hat{a}^{\dag}_{j,k_y}|0\rangle,
\end{equation}
where $\Psi_{j,k_y}$ satisfies \cite{hatsugai1993edge}
\begin{equation}
 -\kappa \left(\Psi_{j+1,k_y} +\Psi_{j-1,k_y}\right) - 2\kappa \cos (k_y -2\pi j\phi_0)
\Psi_{j, k_y} = E_{k_y}\Psi_{j,k_y}
\end{equation}
with $E_{k_y}$ the eigenenergy.

We can now express the photon transmission coefficient in terms of
$|\Psi_{k_y}\rangle$,
\begin{equation}
T_{j,0}^{j_o,l_o}=-i\langle
j_o,l_o|\sum_{\{|\Psi_{k_y}\rangle\}}\left(|\Psi_{k_y}\rangle\frac{\gamma}{\omega-E_{k_y}
+i\gamma/2}\langle\Psi_{k_y}|\right)|j,0\rangle.
\label{eq:T_edge}
\end{equation}
Clearly, only states with energies close to the probing light frequency
$\omega$ have significant contribution to $T_{j,0}^{j_o,l_o}$. Because of this,
when $\omega$ falls in the mid of a gap in the system spectrum and $\gamma$ is
much smaller than the corresponding band gap, we can include only the edge
states in calculating $T_{j,0}^{j_o,l_o}$ in Eq. (\ref{eq:T_edge}) since in-band
states are far off resonance. Deep in the band gap, the dispersion relation of
the edge states is linear in $k_y$ \cite{hatsugai1993edge}. Taking into account the
possibility of multiple edge modes in the vicinity of $\omega$, we have
$E^m_{k_y}=\omega+v_m(k^m_y-\Bbbk_y^m )$, where $v_m$ is the group velocity of
the $m$-th edge mode and $\Bbbk_y^m$ is the Bloch momentum of the state in
resonance with the probing light ($E_{\Bbbk_y^m}^m=\omega$). Making use of the
dispersion relation, we obtain in the continuum limit $N_y\rightarrow\infty$
\begin{equation}
T_{j,0}^{j_o,l_o}\simeq \frac{1}{2\pi}\sum_m\int
dk_y^m\Psi^m_{j_o,k_y}\frac{i\gamma}{(k^m_y-\Bbbk_y^m)v_m
-i\gamma/2}\Psi^{m*}_{j,k_y}e^{ik_y^ml_o},
\label{eq:T_edge_cont}
\end{equation}
where $\Psi^m$ is the $m$-th edge mode. Since only states close to $\Bbbk_y^m$
contribute to the integration in Eq. (\ref{eq:T_edge_cont}), we can evaluate it
by approximating $\Psi^m_{j_o,k_y}$ with $\Psi^m_{j_o,\Bbbk_y^m}$ and extending
the limit of the integration to $(-\infty, \infty)$. The result is
\begin{equation}
T_{j,0}^{j_o,l_o}\simeq \frac{1}{2\pi} \sum_m
\Psi^m_{j_o,\Bbbk_y^m}\Psi^{m*}_{j,\Bbbk_y^m}
\int_{-\infty}^\infty dk_y^m\frac{i\gamma}{(k_y-\Bbbk_y^m)v_m
-i\gamma/2}e^{ik_y^ml_o} =
-\sum_m\Psi^m_{j_o,\Bbbk_y^m}\Psi^{m*}_{j,\Bbbk_y^m}\frac{\gamma}{v_m}
\Theta\big(\frac { l_o } { v_m } \big)e^ {-\frac{\gamma}{2}\frac{l_o}{v_m}}e^{i\Bbbk_y^ml_o}
\label{eq:T_evalated}
\end{equation}
with the step function
\begin{numcases}
{\Theta(x)=} \nonumber
0 & $x<0$\\ \nonumber
\frac{1}{2} & $x=0$\\ \nonumber
1 & $x>0$. \nonumber
\end{numcases}

By using Eq. (\ref{eq:T_evalated}), it is straightforward to calculate the
average OAM number displacement. We obtain
\begin{equation}
 \bar{l}_e=\sum_{j\in \text{edge}}\sum_{j_o,l_o}|T_{j,0}^{j_o, l_o}|^2\cdot
l_o
\simeq\sum_{m \in \text{left}}\text{sgn}(v_m),
\end{equation}
where the summation over $m$ includes only the corresponding edge states on the
left edge of the lattice.
We have used $\sum_{j\in \text{edge}}|\Psi^m_{j,\Bbbk_y^m}|^2 \simeq 1$ when the $m$-th
edge mode is on the left edge and $\sum_{j\in \text{edge}}|\Psi^m_{j,\Bbbk_y^m}|^2 \simeq 0$
when it is on the right edge, which follows from the fact that
the distribution of the edge states is limited to the edge of the lattice.
This result indicates that $\bar{l}_e$ is approximately equal to the difference
between the number of up and down moving edge states, which in turn is equal to
the total Chern number (up to a sign depending on edge transport of the left or right
edge) for the bands below the gap due to the bulk-boundary
correspondence \cite{hasan2010colloquium}.

\section*{Supplementary Note 6: Measurement of the Chern number}
As shown in the main text, the Chern number of a finite lattice can be measured
via the average OAM number displacement ($\bar{l}_e$ in Eq. (\ref{eq:l_e})) in
edge-state transport. For an infinite system, the Chern number is equal to the
TKNN index \cite{thouless1982quantized, hatsugai1993chern,
kohmoto1985topological}. We demonstrate in this section that it can be
calculated from experimentally measured photon transmission coefficients.

The TKNN index in an infinite system is determined by the bulk wave function.
As discussed in Supplementary Note 2, in order to keep the size of the
simulator array small, we should choose a gauge that leads to the Hamiltonian
\begin{equation}
\mathcal{H}_2=-\kappa\sum_{j,l}\big(a_{j,l+1}^{\dagger}a_{j,l}+a_{j,l}^{\dagger}
a_{j,l+1}
 +e^{-i2\pi l\phi_0}a_{j+1,l}^{\dagger}a_{j,l}+e^{i2\pi
 l\phi_0}a_{j,l}^{\dagger}a_{j+1,l}\big),
 \label{eq:H_Abelian_again}
 \end{equation}
where $\phi_0=p/q$ ($p$ and $q$ mutually prime integers) is the flux quanta
per plaquette. The configuration of the simulation system has been described in
Supplementary Note 2.

We use periodic boundary condition in both the $x$ and $y$ directions to
simulate an infinite system.
According to the Bloch theorem, the eigenstates of
$\mathcal{H}_2$ can be written in the form
\begin{equation}
\Psi_{j,l}(k_x,k_y)=e^{ik_yl}e^{ik_xj}u_{l_q}(k_x,k_y),
\end{equation}
where $k_x\in[-\pi,\pi]$, $k_y\in[0,2\pi/q]$ are the Bloch vectors, $l_q=\text{mod}(l,q)
\in [0,q-1]$ is the OAM index within a magnetic unit cell, and
$u_{l_q}(k_x,k_y)=u_{l_q+q}(k_x,k_y)$ is a periodic function.

The spectrum of the system consists of $q$ energy bands \cite{hofstadter1976energy}.
The Chern number (or equivalently the TKNN index) of
the $m$-th ($m \in [1,q]$) band can be expressed as
\cite{thouless1982quantized, hatsugai1993chern,
kohmoto1985topological}
\begin{equation}
C=\frac{1}{2\pi i}\int\int dk_xdk_y(\langle\frac{\partial u^m}{\partial
k_x}|\frac{\partial u^m} {\partial k_y}\rangle-\langle\frac{\partial
u^m}{\partial k_y}|\frac{\partial u^m}{\partial k_x}\rangle)
=\frac{1}{2\pi i}\int\int dk_xdk_y[\nabla_k\times \textbf{A}^m(k_x,k_y)]_z,
\label{eq:C1}
\end{equation}
where $\textbf{A}^m=\langle u^m|\nabla_k|u^m\rangle$ and
\begin{equation}
 |u^m(k_x,k_y)\rangle=[u^m_0(k_x,k_y), \ldots, u^m_{q-1}(k_x,k_y)]^T
\end{equation}
is the eigenstate vector of the $m$-th band. There is a gauge freedom which
comes from the phase ambiguity of $|u^m(k_x,k_y)\rangle$, since
\begin{equation}
e^{if(k_x,k_y)}|u^m(k_x,k_y)\rangle
\label{eq:phase_gauge}
\end{equation}
is also a solution as long as
$f(k_x,k_y)$ is a smooth function of $(k_x,k_y)$ and it is independent of
$(x,y)$. The Chern number is invariant under this gauge transformation.

A non-trivial topology arises when the phase of the wave function cannot be
determined uniquely and smoothly in the entire magnetic Brillouin zone.
In this case, one cannot apply the Stokes theorem globally to evaluate Eq.
(\ref{eq:C1}) \cite{kohmoto1985topological}. Following Refs.
\cite{hatsugai1993chern, kohmoto1985topological}, we divide the Brillouin zone
into two regions B1 and B2 [see Supplementary Figure \ref{fig:ChernS} (b)], where B2 is chosen
such that it contains all zero
points of $u^m_0(k_x,k_y)$
and at least one $u^m_{l_q}(k_x,k_y)$ with $l_q \neq 0$ does not vanish in it.
By taking advantage of the gauge transformation in Eq.
(\ref{eq:phase_gauge}) with an appropriate $f(k_x,k_y)$, we can choose a phase
convention in B1 such that $u^m_{0}(k_x,k_y)$ is real, and another phase
convention in B2 such that $u^m_{l_q}(k_x,k_y)$ is real.
The chosen phase conventions lead to smooth vector fields
$\textbf{A}^m_{B1}$ and $\textbf{A}^m_{B2}$ on B1 and B2 respectively, and
result in a phase mismatch $\chi(k_x,k_y)$ on the boundary of B1 and B2
\cite{kohmoto1985topological},
\begin{equation}
|u^m\rangle_{B1}=e^{i\chi(k_x,k_y)}|u^m\rangle_{B2}.
\label{eq:chi}
\end{equation}
We can then apply Stokes' theorem on B1 and B2 separately to derive
\begin{equation}
C=\frac{1}{2\pi i}\int_{\partial B1}
d\textbf{k}\cdot[\textbf{A}^m_{B1}(k_x,k_y)-\textbf{A}^m_{B2}(k_x,k_y)]
=\frac{1}{2\pi}\int_{\partial B1} d\textbf{k}\cdot \nabla_k \chi(k_x,k_y),
\label{eq:C2}
\end{equation}
where $\partial$B1 is the boundary of B1.

We can obtain $|u^m\rangle$ and determine $\chi(k_x,k_y)$ from photon
transmission measurement and then use Eq. (\ref{eq:C2}) to calculate the Chern
number. Suppose we couple a $l=0$ OAM beam to the first cavity in the
simulator array, which is equivalent to driving the simulated lattice system at
site $(0,0)$, and measure the transmission coefficient to site $(j,l)$,
$T_{0,0}^{{j,l}}$.
The Fourier transformation of $T_{0,0}^{{j,l}}$  to the momentum space
$(k_x,k_y)$, $T(k_x,k_y,l_q)\propto \sum_{j,l}
T_{0,0}^{(j,ql+l_q)}e^{-ik_xj}e^{-ik_y(ql+l_q)}$,
is given by
\begin{equation}
T(k_x,k_y,l_q)\propto \langle
k_x,k_y,l_q|\frac{i\gamma}{\omega-\mathcal{H}+i\gamma}|j=0,l=0\rangle
\end{equation}
where $|k_x,k_y,l_q\rangle\propto \sum_{j,l}e^{ik_xj}e^{ik_y(ql+l_q)}|j,ql+l_q\rangle$.
If the photon loss rate $\gamma$ is much smaller than
the band gaps, and the driving frequency is close to the $m$-th band, only
states in the $m$-th band are excited and contribute to the transmission.
Consequently,
\begin{equation}
T(k_x,k_y,l_q)\propto  u^m_{l_q}(k_x,k_y)
\frac{i\gamma}{\omega-E_m(k_x,k_y)+i\gamma}u^{m}_{0}(k_x,k_y)^*,
\end{equation}
where $E_m(k_x,k_y)$ is the energy of the $m$-th band at $(k_x,k_y)$. By
using a similar idea in \cite{PhysRevLett.112.133902}, for each
$(k_x,k_y)$ we can fine tune the driving frequency such that it is in resonance
with $E_m(k_x,k_y)$, i.e. $\omega-E_m(k_x,k_y)\ll \gamma$.
This then allows us to relate the photon transmission coefficient to the
wave function in the $m$-th band via
\begin{equation}
T(k_x,k_y,l_q)\propto  u^m_{l_q}(k_x,k_y)u^{m}_{0}(k_x,k_y)^*.
\label{eq:T_u}
\end{equation}

By using Eq. (\ref{eq:T_u}) and renormalizing the measured $T(k_x,k_y,l_q)$, we
can determine the eigenstate
$|u^m(k_x,k_y)\rangle=[u^m_0(k_x,k_y), u^m_1(k_x,k_y), \ldots,
u^m_{q-1}(k_x,k_y)]^T$ of
the $m$-th band. With the help of the gauge transformation in Eq.
(\ref{eq:phase_gauge}), we can further choose the phase of the eigenstate $|u^m\rangle$ in the magnetic
Brillouin zone using the technique discussed earlier.
This then allows us to determine
$\chi(k_x,k_y)$ in Eq. (\ref{eq:chi}) and calculate the Chern number according
to Eq. (\ref{eq:C2}).

As an example, we consider the flux $p/q=1/6$, and show how to measure
the Chern number of the first band ($m=1$). From the band structure in Supplementary Figure
\ref{fig:ChernS} (a), we see that this band is located near $\omega=-3.09\kappa$
and it is very narrow. With a photon loss rate of $\gamma=0.1\kappa$, which is
much larger than the width of this band and much smaller than the band gaps
surrounding it, we can achieve resonance with all states in it and avoid exciting
states in other bands by fixing the frequency of the probing light at
$\omega=-3.09\kappa$.

We then divide the magnetic Brillouin zone into two areas as prescribed earlier.
Specifically, we define $B1=\{k_x \in [-0.4\pi,0.4\pi], ky \in [0,2\pi/q]\}$,
and the rest B2, as depicted in Supplementary Figure \ref{fig:ChernS} (b). In B1,
$u^1_0(k_x,k_y)$ is always nonzero. B2 contains all the zero points of
$u^1_0(k_x,k_y)$. Also, $u^1_3$ does not vanish in B2. As discussed earlier,
with this division we can define two different phase conventions for the
eigenstates in B1 and B2 \cite{hatsugai1993chern, kohmoto1985topological}. In
one convention,
$u^1_0(k_x,k_y)$ is real in B1. In the other convention, $u^1_3(k_x,k_y)$ is
real in B2.
From Eq. (\ref{eq:chi}), we see that the phase
mismatch $\chi(k_x,k_y)$ on the boundary $\partial$B1 is given by the phase of
$u^1_3(k_x,k_y)$ on $\partial$B1. According to Eq. (\ref{eq:T_u}), if we
drive the simulated system at site $(0,0)$, we have $T(k_x,k_y,l_q)\propto
u^1_{l_q}(k_x,k_y)u^{1}_{0}(k_x,k_y)^*$, from which we can obtain
$|u^1\rangle\propto[T(k_x,k_y,0), T(k_x,k_y,1), \ldots, T(k_x,k_y,5)]^T$.
Therefore, $\chi(k_x,k_y)$ is given by the phase of
$T(k_x,k_y,3)$ relative to that of $T(k_x,k_y,0)$ on $\partial$B1,
boundary of B1,
and the Chern number can be calculated using Eq. (\ref{eq:C2}).

\section*{Supplementary References}